\documentclass[11pt]{article}
\usepackage[margin=1in]{geometry}
\usepackage{amsmath,amsthm,amssymb}
\usepackage{hyperref}
\usepackage{authblk}

\usepackage{bm}
\usepackage{bbm}

\usepackage{pgf}
\usepackage{tikz}
\usetikzlibrary{arrows,automata}

\newcommand{\removeforcameraready}[1]{}

\newcommand{\adv}{\text{\rm adv}}

\newcommand{\numsamples}{t}
\newcommand{\clusterrate}{\ell}

\newcommand{\eps}{{\varepsilon}}

\newcommand{\indicator}[1]{\mathbbm{1}\{#1\}}

\newcommand{\outputspace}{\mathcal{O}}
\newcommand{\LCS}{d_{\text{LCS}}}

\newcommand{\E}{\mathbb{E}}
\newcommand{\Lap}{\operatorname{Lap}}

\newcommand{\neymanalloc}{{\bm{f}_{\rm Neyman}}}

\newcommand{\weightedmean}{\theta_{\mu}}

\newcommand{\argmin}[1]{\underset{#1}{\operatorname{arg}\,\operatorname{min}}\;}

\newcommand{\datauni}{\mathcal{U}}
\newtheorem{thm}{Theorem}[section]
\newtheorem{theorem}[thm]{Theorem}

\newtheorem{lemma}[thm]{Lemma}

\newtheorem{proposition}[thm]{Proposition}

\newtheorem{definition}[thm]{Definition}

\title{Controlling Privacy Loss in Sampling Schemes:
an Analysis of Stratified and Cluster Sampling}

\author[1]{Mark Bun\thanks{{Supported also by NSF grants CCF-1947889 and CNS-2046425.}}}

\author[2,3]{J\"org Drechsler}

\author[1]{Marco Gaboardi}

\author[4]{Audra McMillan\thanks{Part of this work was done while AM was supported by a Fellowship from the Cybersecurity \& Privacy Institute at Northeastern University and NSF grant CCF-1750640.}}

\author[5]{Jayshree Sarathy\thanks{Corresponding author}}

\affil[1]{Department of Computer Science at Boston University, USA}
\affil[2]{Institute for Employment Research, Germany}
\affil[3]{Joint Program in Survey Methodology, University of Maryland, USA}
\affil[4]{Apple, USA}
\affil[5]{Harvard John A. Paulson School of Engineering and Applied Sciences, USA}

\date{}

\begin{document}

\maketitle

\begin{abstract}
Sampling schemes are fundamental tools in statistics, survey design, and algorithm design. 
A fundamental result in differential privacy is that a differentially private mechanism run on a \emph{simple random} sample of a population provides stronger privacy guarantees than the same algorithm run on the entire population. 
However, in practice, sampling designs are often more complex than the simple, data-independent sampling schemes that are addressed in prior work.
In this work, we extend the study of privacy amplification results to more complex, data-dependent sampling schemes. We find that not only do these sampling schemes often fail to amplify privacy, they can actually result in privacy degradation.
We analyze the privacy implications of the pervasive cluster sampling and stratified sampling paradigms, as well as provide some insight into the study of more general sampling designs.
\end{abstract}

\section{Introduction}

Sampling schemes are fundamental tools in statistics, survey design, and algorithm design.   
For example, they are used in social science research to conduct surveys on a random sample of a target population. They are also used in machine learning to improve the efficiency and accuracy of algorithms on large datasets.
In many of these applications, however, the datasets are sensitive and privacy is a concern.
Intuition suggests that (sub)sampling a dataset before analysing it provides additional privacy, since it gives individuals plausible deniability about whether their data was included or not. 
This intuition has been formalized for some types of sampling schemes (such as simple random sampling with and without replacement and Poisson sampling) in a series of papers in the differential privacy literature~\cite{KasiviswanathanLNRS11,WangLF16b,BunNSV15, WangBK19}. Such \emph{privacy amplification by subsampling} results 
can provide tight privacy accounting when analysing algorithms that incorporate subsampling, e.g.,~\cite{WangFS15,AbadiCGMMT016,JalkoHD17,ParkFCW20,GirgisDDKS21}. However, in practice, sampling designs are often more complex than the simple, data independent sampling schemes that are addressed in prior work. In this work, we extend the study of privacy amplification results to more complex and data dependent sampling schemes. 

\begin{figure}
\centering
\begin{tikzpicture}[->,>=stealth,shorten >=2pt,auto,node distance=4cm,
scale = 1,transform shape,very thick]

\node (q_0) {$P, H$};
\node (q_1)  [right of=q_0]  {$P, \mathcal{C}(H)$};
\node(q_2) [right of=q_1] {$S$};
\node (q_3) [right of=q_2] {$\mathcal{M}(S) = \mathcal{M}_\mathcal{C}(H, P)$};

\path 
(q_0) edge              node[scale=0.7, align=center] {Historic or auxiliary data \\ is used to design \\ the sampling scheme} (q_1)
(q_0) edge    [bend right=20]        node[scale=0.7, align=center] {End-to-end process \\ described by $\mathcal{M}_\mathcal{C}$} (q_3)
(q_1) edge             node[scale=0.7, align=center] {Sampling scheme $\mathcal{C}(H)$ is \\ used to sample dataset $S$ \\ from population $P$} (q_2)
(q_2) edge             node[scale=0.7, align=center] {$\eps$-differentially private \\ mechanism $\mathcal{M}$ \\ is run on input $S$} (q_3)
;
\end{tikzpicture}

\caption{The structure of using a data-dependent sampling scheme.}
\label{fig:sampling-desc}
\end{figure}

We consider the setting described in Figure~\ref{fig:sampling-desc}. We have a \emph{population} $P$ and a historic or auxiliary data set $H$ which is used to inform the sampling design. We think about the sampling scheme as a function $\mathcal{C}(H)$ of the historic or auxiliary data $H$. Using this sampling scheme, we draw a sample $S$ from the population $P$, on which we run the differentially private mechanism $\mathcal{M}$. We can think about these multiple steps as comprising a mechanism $\mathcal{M}_{\mathcal{C}}(H,P)$ working directly on the population $P$ and the historic data $H$ whose privacy depends on both the mechanism $\mathcal{M}$ and the sampling scheme $\mathcal{C}(H)$. While this is the general framework for the problem we study, we state the technical results in this paper for the simplified case where $H = P$; see Section~\ref{sec:setting} for further discussion.

\subsection{Our contributions} 
We primarily focus on three classes of sampling schemes that are common in practice: \emph{cluster sampling}, \emph{stratified sampling}, and \emph{sampling with probability proportional to size (PPS)}. In (single-stage) cluster sampling, the population arrives partitioned into disjoint clusters. A sample is obtained by selecting a small number of clusters at random, and then including all of the individuals from those chosen clusters. In stratified sampling, the population is partitioned into ``strata.'' Individuals are then sampled independently from each stratum, potentially with different sampling rates between the strata. In PPS sampling, the probability of selecting each individual depends on some measure of size available for all individuals in the population.

For these more complex schemes, we find that privacy amplification can be negligible even when only a small fraction of the population is included in the final sample. Moreover, in settings where the sampling design is data dependent, privacy degradation can occur -- some sampling designs can actually make privacy guarantees worse. Intuitively, this is because the sample design itself can reveal sensitive information.
Our goal in this paper is to explain how and why these phenomena occur and introduce technical tools for understanding the privacy implications of concrete sampling designs.

\smallskip

\noindent \textbf{Understanding randomised and data-dependent sampling.} It is simple to show that deterministic, data-dependent sampling designs do not achieve privacy amplification, and can suffer privacy degradation. Motivated by this observation, we start by studying the privacy implications of randomised and data-dependent sampling, attempting to isolate their effects in the simplest possible setting.

Specifically, we aim to understand sampling schemes of the following form: For a possibly randomised function $f$ (an ``allocation rule''), sample $f(P)$ individuals uniformly from $P$ without replacement. In Section~\ref{RR}, we study the case where $f$ is randomised but data-independent, i.e., the number of individuals to be sampled is drawn from a distribution that does not depend on $P$. We give an essentially complete characterization of what level of amplification is possible in terms of this distribution.

In Section~\ref{datadependentsampling}, we turn our attention to data-dependent sampling. We identify necessary conditions for an allocation rule $f$ to enable privacy amplification by way of a hypothesis testing perspective; intuitively, for $f$ to be a good amplifier, every differentially private algorithm must fail to distinguish the distributions of $f(P)$ and $f(P')$ for neighboring $P, P'$. We also study a specific natural allocation rule called \emph{proportional allocation} that is commonly applied in stratified sampling. We design a simple randomised rounding method that offers a minor change to the way proportional allocation is generally implemented in practice, but that offers substantially better privacy amplification. 

\smallskip

\noindent\textbf{Cluster sampling.} In Sections~\ref{clustersampling} and~\ref{othersamplingschemes}, we turn to more complex sampling designs, including cluster sampling. 
In Section~\ref{clustersampling}, we study cluster sampling with simple random sampling of clusters. In this sampling design, a population partitioned into $k$ clusters is sampled by selecting $m$ clusters uniformly at random without replacement. Our results give trade-offs between the privacy amplification achievable and the sizes of the clusters. In particular, privacy amplification is possible when all of the clusters are small. As the cluster sizes grow, the best achievable privacy loss rapidly approaches the baseline privacy guarantee (i.e., the privacy guarantee we would get without any sampling). 
We provide some insight into these results by connecting the privacy loss to the ability of a hypothesis test to determine from a differentially private output which clusters were included in the sample. 

In Section~\ref{othersamplingschemes}, we then discuss other common sampling schemes like Probability-Proportional-to-Size (PPS) sampling and systematic sampling. We attempt to identify properties of sampling schemes that are red flags, indicating that a sampling scheme may not provide a general amplification result. 

\smallskip
\noindent \textbf{Stratified sampling.} 
In Section~\ref{sec:stratified}, we turn our attention to stratified sampling. In stratified sampling, the population is stratified into sub-populations and sampling is performed independently within each strata. Much the intuition developed in the single stratum case also holds in the case of multiple strata. An allocation rule $f$ is often used to decide how many samples to draw from each stratum.
A common goal when choosing an allocation function $f$ is to minimise the variance of a particular statistic. For example, the popular Neyman allocation is the optimal allocation for computing the population mean. A natural question then is how to define and compute the optimal allocation when privacy is a concern? 
In this work, we will formulate the notion of an optimal allocation under privacy constraints. This formulation is somewhat subtle since the privacy implications of different allocation methods need to be properly accounted for. 
Our goal is to initiate the study of alternative allocation functions that may prove useful when privacy is a concern. 

Finally, building on our randomised rounding method for the ``single-stratum'' case, we show that stratified sampling with the common \emph{proportional allocation rule} does indeed amplify privacy. 

\subsection{Related work} 
Several works have studied the privacy amplification of simple sampling schemes. 
Kasiviswanathan et al.~\cite{KasiviswanathanLNRS11} and Beimel et al.~\cite{BeimelKN10} showed that applying Poisson sampling before running a differentially private mechanism improves its end-to-end privacy guarantee. Subsequently, Bun et al.~\cite{BunNSV15} analyzed simple random sampling with replacement in a similar way.
Beimel et al.~\cite{BeimelNS13}, 
Bassily et al.~\cite{BassilyST14}, and 
Wang et al.~\cite{WangLF16} analyzed simple random sampling without replacement. Imola and Chaudhuri~\cite{ImolaC21} provide lower and upper bounds on privacy amplification when sampling from a multidimensional Bernoulli family, a task which has direct applications to Bayesian inference.
Balle et al.~\cite{BalleBG18} unified the analyses of privacy amplification of these mechanisms using the lenses of \emph{probabilistic couplings}, an approach that we also use in this paper. The effects that sampling can have on differentially private mechanisms is also studied from a different perspective in~\cite{EbadiAS16}. However, none of the prior works consider the privacy amplification of more complex, data-dependent sampling schemes
commonly used in practice. To the best of our knowledge, this paper is the first to do so.
 
\section{Background}

\subsection{Data-dependent sampling schemes} \label{sec:setting}

In the data-driven sciences, data is often obtained by sampling a subset of the population of interest. This sample can be created in a wide variety of ways, referred to as the sample design. Sample designs can vary from simple designs such as taking a uniformly random subset of a fixed size, to more complex data-dependent sampling designs like cluster or stratified sampling. Data-dependent sampling designs balance accuracy and financial requirements by using historic or auxiliary data to exploit structure in the population. The privacy implications of simple random sampling are quite well understood from prior work. In this work, we will move beyond simple random sampling to analyse the privacy implications of more complex sampling designs, including data-dependent sampling.

An outline of the schema for data dependent sampling designs is given in Figure~\ref{fig:sampling-desc}. There are two datasets: $H$, the historic or auxiliary data that is used to design the sampling scheme $\mathcal{C}(H)$, and $P$, the current population that is sampled from. 
For the remainder of this paper, we make the simplifying assumption that $H=P$. That is, we will not distinguish between the historic or auxiliary data and the ``current'' data. Of course, this is not strictly true since these datasets serve different purposes. However, $H$ and $P$ are likely \emph{correlated}. For example, they may contain information about the same set of data subjects. Since we typically will not understand the extent of the correlation, if we want to protect the information in the dataset $P$, we also need to protect the information in the dataset $H$. From this perspective, we do not lose anything by assuming the $P=H$ from the outset.
Thus, we view the function $\mathcal{M}_{\mathcal{C}}(P,H)$ as simply a function of $P$. 
More refined models may be obtained by imposing specific assumptions on the relationship between $H$ and $P$, for example, by modeling the temporal correlation between historic and current data.
We leave this direction for future work.

Throughout this paper, we will refer to the size of the sample $S$ as the \emph{sample size}, and the fraction $|S|/|P|$ as the \emph{sampling rate}. Furthermore, we will assume that the mechanism $\mathcal{M}$ only has access to the sample itself, not the specifics of the sampling design. For example, this precludes mechanisms that re-weight samples according to their probability of selection (unless this probability is data independent). In the final section, we will discuss some of the challenges associated with incorporating sampling weights under privacy. This is an interesting and important direction for future research, although out of scope for this paper.

\subsection{Differential privacy}

Differential privacy (DP) is a measure of stability for randomised algorithms. It bounds the change in the distribution of the outputs of a randomised algorithm when provided with two datasets differing on the data of a single individual. We will call such datasets neighboring. 
In order to formalise what a ``bounded change'' means, we define $(\eps, \delta)$-indistinguishability. Two random variables $P$ and $Q$ over the same probability space are $(\eps, \delta)$-\emph{indistinguishable} if for all sets of outcomes $E$ over that probability space, 
\[e^{-\eps}(\Pr(Q\in E)-\delta) \le \Pr(P\in E)\le e^{\eps}\Pr(Q\in E)+\delta.\]
If $\delta=0$ then we will say that $P$ and $Q$ are $\eps$-indistinguishable. 
For any $n\in\mathbb{N}$, let $\datauni^n$ be the set of all datasets of size $n$ over elements of the data universe $\datauni$.
Let $\datauni^{*}=\cup_{n\in\mathbb{N}} \datauni^{n}$ be the set of all possible datasets. We discuss two privacy definitions in this work corresponding to two different neighboring relations: \emph{deletion} differential privacy and \emph{replacement} differential privacy. 
We will say two datasets are \emph{deletion neighbors} if one can be obtained from the other by adding or removing a single data point, 
and \emph{replacement} neighbors if they have the same size, and one can be obtained from the other by changing the data of a single individual.

\begin{definition}\label{DP}
A mechanism $\mathcal{M}:\datauni^*\to\outputspace$ is $(\eps, \delta)$-deletion (resp. replacement) differentially private (DP) if for all pairs of deletion (resp. replacement) neighboring datasets $P$ and $P'$, $\mathcal{M}(P)$ and $\mathcal{M}(P')$ are $(\eps,\delta)$-indistinguishable.
\end{definition}

We will use both replacement and deletion DP throughout the paper as they are appropriate in different settings. When considering which notion to choose, it is important to consider which guarantees are meaningful in context. For example, it will be common in the sample designs we cover for the size of the sample $S$ (see Figure~\ref{fig:sampling-desc}) to be data-dependent. When considering these sampling designs, we will focus on mechanisms $\mathcal{M}$ that satisfy deletion DP since replacement DP does not protect the sample size. 
However, replacement DP may be more appropriate for the privacy guarantee on $\mathcal{M}_{\mathcal{C}}$ in applications where it is unrealistic to assume that an individual can choose not to be part of the auxiliary dataset or the population. For example, the auxiliary data may be administrative data, data from a mandatory census, or data from a monopolistic service provider. Results and intuition are often similar between deletion and replacement DP, although care should be taken when translating between the two notions. We note in particular that any $\eps$-deletion DP mechanism is $2\eps$-replacement DP.

\subsection{Privacy amplification with uniform random sampling} 

Sampling does not provide strong differential privacy guarantees on its own.
But when employed as a pre-processing step in a differentially private algorithm, it can amplify existing privacy guarantees.
Intuitively, this is because if the choice of individuals is kept secret, sampling provides data subjects the plausible deniability to claim that their data was or was not in the final data set. 
This effect was first explicitly articulated in~\cite{Smith10, Ullman:2017}, and a formal treatment of the phenomenon was given in \cite{BalleBG18}. Three types of sampling are analysed in \cite{BalleBG18}: simple random sampling with replacement, simple random sampling without replacement, and Poisson sampling. In all three settings the privacy amplification is proportional to the probability of an individual not being included in the final computation. 
To gain some intuition before we move into the more complicated sampling schemes that are the focus on this paper, let us state and discuss the results from \cite{BalleBG18}.

\begin{theorem}\label{priorwork}\cite{BalleBG18}
Let $\mathcal{C}$ be a sampling scheme that uniformly randomly samples $m$ values out of $n$ possible values without replacement. Given an $(\eps,\delta)$-replacement differentially private mechanism $\mathcal{M}$, we have that $\mathcal{M}_\mathcal{C}$ is $(\eps',\delta')$-replacement differentially private for $\eps'=\log(1+\frac{m}{n}(e^\eps -1))$ and 
$\delta'=\frac{m}{n}\delta$.
\end{theorem}

To consider the implications of this result, notice that $\eps'\le\eps$ for all values of $m\le n$ so the sampled mechanism $\mathcal{M}_{\mathcal{C}}$ is strictly more private than the original mechanism $\mathcal{M}$. Further, taking into account the following two approximations which hold for small $x$, 
\begin{align}
    e^x-1 &\approx x \label{approximation-one} \\
    \log(1+x) &\approx x, \label{approximation-two}
\end{align} 
we have that for small $\eps$, $\eps'\approx \frac{m}{n}\eps$. So the degree of amplification in both parameters is roughly proportional to the sampling rate $m/n$.

\subsection{How do people use subsampling amplification results?}\label{howtouse}

Suppose we have a dataset that contains $n$ records, and we want to estimate the proportion of individuals that satisfy some attribute in an $\eps$-DP manner.
Let us set our target privacy guarantee to be $\eps=1$.
To do this, we can simply compute the proportion non-privately and add Laplace noise with scale $1/n$. But, if we know that the dataset is a secret and simple random sample from a population of $100n$ individuals,
 then adding Laplace noise with scale $1/n$ as before will actually yield a stronger privacy guarantee of $\eps' = 0.01$ for the underlying population. To get $\eps' = 1$, we will need to add noise with scale only $1/(100n)$. In other words, the secrecy of the sample means that the computation has more privacy inherently, and therefore, we can add less noise in order to achieve the desired privacy guarantee.

Existing DP data analysis tools such as DP Creator~\cite{GaboardiHKMNUV16,GaboardiHV20}
employ privacy amplification results to provide better statistical utility. For example, the DP Creator interface prompts the user to input the population size if the data is a secret and random sample from a larger population of known size and take advantage of the resulting boost in accuracy without changing the privacy guarantee.

As we discussed before,  privacy amplification results are also used to  analyse algorithms that incorporate subsampling as one of their components. Privacy amplification results permit a tighter analysis of the privacy that these algorithm can guarantee.  In particular, these algorithms are quite common in learning tasks, e.g.~\cite{WangFS15,AbadiCGMMT016,JalkoHD17,ParkFCW20,GirgisDDKS21}.

\section{Randomised data-independent sampling rates}\label{RR}

While we are ultimately interested in data-dependent sampling designs, we begin with an extension of Theorem~\ref{priorwork} to data-\emph{independent}, but randomised, sampling rates. Prior results on privacy amplification by subsampling \cite{KasiviswanathanLNRS11,WangLF16b,BunNSV15, WangBK19, Balle:2018} all focus on constant sampling rates where the sampling rate (the fraction of the data set sampled) is fixed in advance. However, we will later see that randomising the sample rate is essential to privacy amplification when  the target rate is data dependent. To work toward this eventual discussion, we first study the data-independent case to gain intuition for what properties of the distribution on sampling rates characterize how much privacy amplification is possible.

Suppose that there is a random variable $\numsamples$
on $[n]$ such that the allocation rule is defined as $f(P)=\numsamples$ for all databases $P$. That is, the sampling scheme is as follows: given a dataset $P$, a sample $m$ is drawn from $\numsamples$, and then $m$ subjects are drawn without replacement from $P$ to form the sample $S$. In this section we consider deletion differential privacy\footnote{Note that we must use the deletion differential privacy definition for $\mathcal{M}$ in this setting; otherwise, the sample size $m$ could be released in the clear as part of the output of $\mathcal{M}$.} for $\mathcal{M}$ and replacement differential privacy for $\mathcal{M}_{\mathcal{C}}$, where the total number of cases, $n$, is known and fixed. 
A simple generalisation of Theorem~\ref{priorwork} immediately implies that the privacy loss of this randomised scheme is no worse than if $\numsamples$ was concentrated on the maximum value in its support. However, prior work does not give insight into what happens when $\numsamples$ has significant mass below its maximum. What property of the distribution characterises its potential for privacy amplification?
The following theorem characterizes the privacy amplification of sampling without replacement with data-independent randomised sampling rates. 

\begin{theorem}
\label{thm:random-data-independent-sampling}
Let $P$ be a dataset of size $n$, let $\numsamples$ be a distribution over $\{0, 1, \ldots, n\}$, and let $\mathcal{C}: \mathcal{X} \rightarrow \mathcal{U}^*$ be the randomised, dataset-independent sampling scheme that
randomly draws $m \sim \numsamples$ and samples $m$ records from $P$ without replacement. Define the distribution $\tilde{\numsamples}$ on $[n]$ where $\tilde{\numsamples}(m) \propto e^{\eps m} \cdot \numsamples(m)$ for all $m \in [n]$.\\
\textbf{Upper bound:}
Let $\mathcal{M}: \mathcal{U}^* \rightarrow \mathcal{O}$ be an $\eps$-deletion DP algorithm. Then, $\mathcal{M}_{ \mathcal{C}}$ is $\eps'$-replacement DP, where 
\[
\eps' = \log \left(1 + \frac{1}{n} \cdot \mathbb{E}_{m \sim \tilde{\numsamples}}[m] \cdot (e^{\eps} - 1) \right).
\]
\textbf{Lower bound:} There exist neighboring datasets $P$ and $P'$ of size $n$, and an $\eps$-deletion DP mechanism $\mathcal{M}$ such that if $\mathcal{M}_{\mathcal{C}}(P)$ and $\mathcal{M}_{\mathcal{C}}(P')$ are $\eps'$-indistinguishable then \[\eps'\ge -\log \left(1 - \frac{1}{n} \cdot \mathbb{E}_{m \sim \tilde{\numsamples}}[m] \cdot (1-e^{-\eps}) \right)\]
\end{theorem}

The proof of Theorem~\ref{thm:random-data-independent-sampling} appears in Appendix~\ref{sec:randomized-data-independent}. First notice that  Theorem~\ref{thm:random-data-independent-sampling} comports with the generalization of Theorem~\ref{priorwork};
as expected, if the support of $\numsamples$ is contained within $[0,m']$ then $\mathbb{E}_{m \sim \tilde{\numsamples}}[m]\le m'$, so the randomised scheme is at least as private as if $\numsamples$ was concentrated on $m'$.
It also determines that the property of $\numsamples$ that determines the privacy amplification is $\mathbb{E}_{m \sim \tilde{\numsamples}}[m]$, the expectation of an exponential re-weighting of the distribution that gives more weight to larger sample sizes.  
When $\eps$ is small, the simple approximations $e^x-1\approx x$, $1-e^{-x}\approx x$, and $\log(1+x)\approx x$ mean that both the upper and lower bounds amount to \[\eps'\approx\frac{\mathbb{E}_{m \sim \tilde{\numsamples}}[m]}{n}\cdot\eps.\]
Due to the exponential re-weighting,
\[\mathbb{E}_{m \sim \tilde{\numsamples}}[m] = \frac{\sum_{m=0}^n e^{\eps m}\Pr(\numsamples=m)m}{\sum_{m=0}^n e^{\eps m}\Pr(\numsamples=m)}\]
rapidly approaches $n$ as the weight of $\numsamples$ on values close to $n$ increases.  Intuitively, this means that even a small probability of sampling the entire dataset can be enough to ensure that there is no privacy amplification, even if the mode of $\numsamples$ is much smaller than $n$. Conversely, if $\numsamples$ is a light tailed distribution (say, subgaussian) concentrated on a value much smaller than $n$, then privacy amplification is possible.

\begin{figure*}
    \centering
        \begin{minipage}[b]{\textwidth}
        \centering
        \includegraphics[width=.6\textwidth]{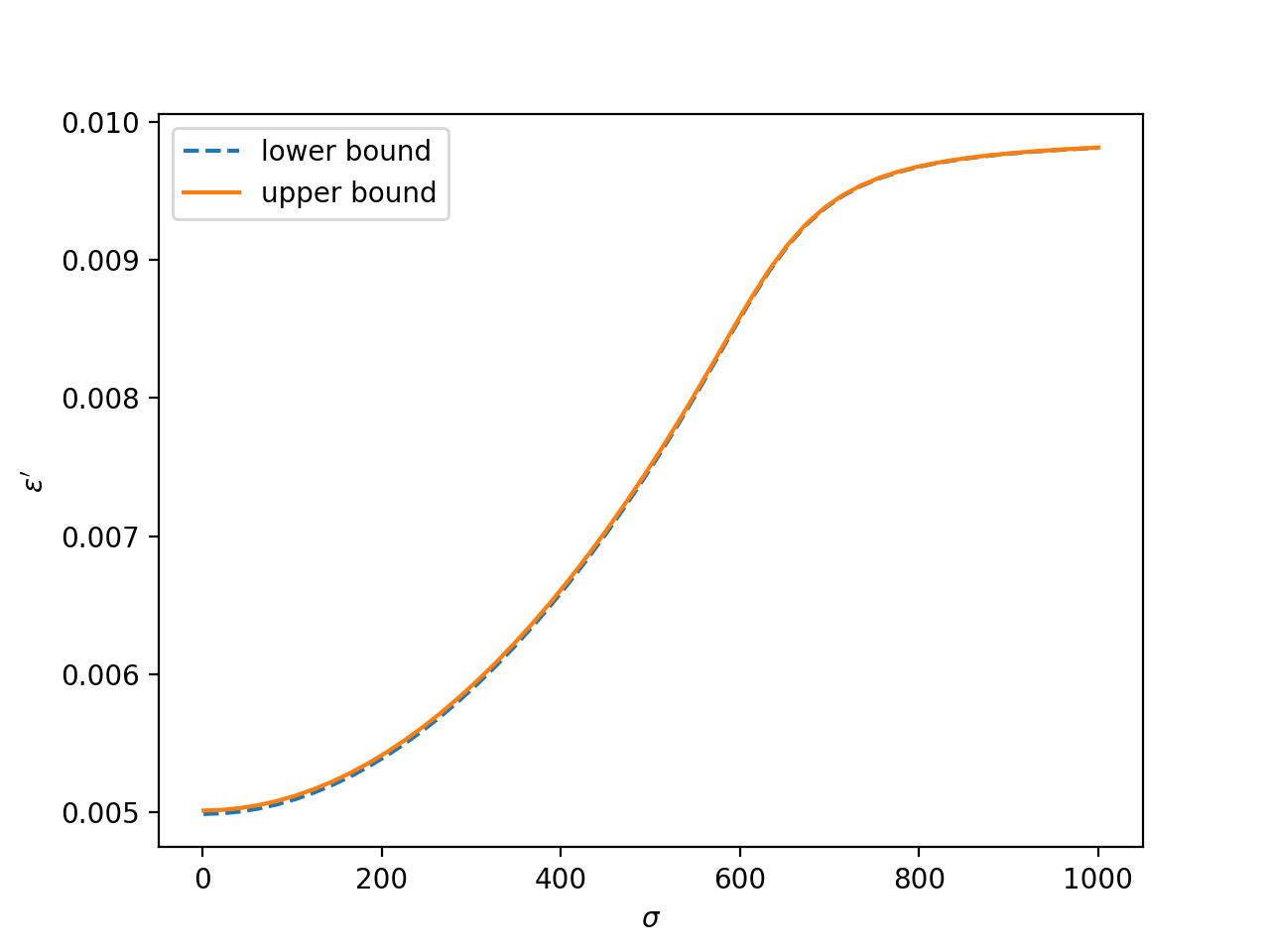}
        \caption{Numerical computation of the upper and lower bounds from Theorem~\ref{thm:random-data-independent-sampling} when $\numsamples$ is truncated Gaussian supported on $[0,n]$ with mean $n/2$, where $n=10^4$ and standard deviation $\sigma$ varies from $1$ to $10^3$. The privacy parameter of the mechanism $\mathcal{M}$ is $0.01$. 
        }
            \label{fig:data-independent}
    \end{minipage}
\end{figure*}

For example, suppose that $\numsamples$ is a truncated Gaussian on $[0,n]$ with mean $n/2$ and standard deviation $\sigma$. If $\numsamples$ is highly concentrated then we expect the privacy guarantee of $\mathcal{M}_{\mathcal{C}}$ to be $\approx\eps/2$. As $\sigma$ grows we expect the privacy guarantee to tend towards $\eps$ as more weight is placed near $n$. In Figure~\ref{fig:data-independent},
we illustrate the bounds of Theorem~\ref{thm:random-data-independent-sampling} numerically with this Gaussian example. We can see that when $n = 10,000$ and $\sigma\approx 800$, the privacy guarantee of $\mathcal{M}_{\mathcal{C}}$ is already close to $\eps = 0.01$, the privacy guarantee of $\mathcal{M}$.

\section{Data-dependent sampling rates}\label{datadependentsampling}

We now turn our attention to sampling schemes where sampling rates may depend on the data. The results in this section are motivated by \emph{stratified sampling}, where the population is stratified into $k$ disjoint sub-populations called strata, and an allocation function is used to determine how many samples to draw from each stratum.
We will discuss stratified sampling with $k > 1$ in Section~\ref{sec:stratified}, but
for simplicity and clarity, we first focus on the ``single stratum'' case. In this section, we develop tools and intuition that we expect to be more broadly useful in understanding complex sampling designs.

Specifically, we consider the sampling design where one selects a number of cases according to a data-dependent function, and then samples that many cases via simple random sampling. That is, let $f : \datauni^* \to \mathbb{N}$ 
be a possibly randomised function
and let $\mathcal{C}_{f}$ be the sampling function that on input $P$ samples $f(P)$ data points uniformly without replacement from $P$.
The question then is, if $\mathcal{M}$ is an $\eps$-DP algorithm, then how private is $\mathcal{M}_{\mathcal{C}_f}$?

\subsection{Sensitivity and privacy degradation}\label{obviousdegradation}

We first observe that if the function $f$ used to determine sample size is highly sensitive, then privacy \emph{degradation} may occur. That is, if the number of cases sampled can change dramatically on neighboring populations, then the output of a DP mechanism can immediately be used to distinguish between those populations. For example, suppose $P$ and $P'$ are neighboring populations, and $f$ is a function where $f(P) = m$ and $f(P') = m + \Delta$. Consider the $\eps$-DP algorithm $\mathcal{M}^{\text{count}}$ that, on input a sample $S$, outputs the noisy count $|S| + \Lap(1/\eps)$ of the number of cases in the sample. Then $\mathcal{M}^{\text{count}}_{ \mathcal{C}_{f}}(P)$ is distributed as $m + \Lap(1/\eps)$ whereas $\mathcal{M}^{\text{count}}_{ \mathcal{C}_{f}}(P')$ is distributed as $m + \Delta + \Lap(1/\eps)$. When $\Delta \gg 1$, these distributions are far apart; the privacy loss between these two populations is $\Delta \cdot \eps \gg \eps$.

Thus, a \emph{necessary} condition for achieving privacy amplification (rather than degradation) is that the function $f$ has low sensitivity or some degree of randomization. In the following sections, we explore other conditions on low sensitivity functions that are necessary and sufficient for amplification.

\subsection{Data dependent sampling and hypothesis testing}

We established in the previous section that using a deterministic data-dependent function to determine sample size results in privacy degradation. This raises the question: how much randomness is necessary to ensure privacy control? 
That is, what can we say about a randomised function $\tilde{f}:\datauni^*\to\mathbb{N}$ with the property that  $\mathcal{M}_{ \mathcal{C}_{f}}$ is $\eps'$-DP for every $\eps$-DP mechanism $\mathcal{M}$? In this section we establish a connection between the amplification properties of a function $\tilde{f}$ and hypothesis testing.

A simple hypothesis testing problem is specified by two distributions $X$ and $Y$. A hypothesis test $H$ for this problem attempts to determine whether the samples given as input are drawn i.i.d from $X$ or from $Y$. If a hypothesis test is only given a single sample then we define the advantage of $H$ to be \[\adv(H; X,Y)=\Pr_{m\sim X}[H(m)=X]-\Pr_{m\sim Y}[H(m)=X].\]
That is, the advantage is a measure of how likely the hypothesis test $H$ is to correctly guess which distribution the sample was drawn from. The closer the advantage is to 1, the better the test is at distinguishing $X$ from $Y$.

One common intuitive explanation of differential privacy is that an algorithm is differentially private if it is impossible to confidently guess from the output which of two neighbouring datasets was the input dataset. This interpretation can be formalised, following~\cite{WassermanZ10}, by noting that if $\mathcal{M}$ is $\eps$-DP and $P$ and $P'$ are neighbouring populations then for every hypothesis test $H$, \[\adv(H; \mathcal{M}(P), \mathcal{M}(P'))\le e^{\eps}-1\approx \eps,\] where the approximation only holds when $\eps$ is small. 

We can establish a similar bound and interpretation of what it means for $\tilde{f}$ to amplify or preserve privacy. Suppose that $\tilde{f}$ is such that $\mathcal{M}_{\mathcal{C}_{\tilde{f}}}$ is $\eps'$-DP for every $\eps$-DP mechanism $\mathcal{M}$. Then in particular, for every $\eps$-DP hypothesis test $H$, we have that $H_{\mathcal{C}_{\tilde{f}}}(P)$ and $H_{\mathcal{C}_{\tilde{f}}}(P')$ are $\eps'$-indistinguishable. Now, if we consider only hypothesis tests $H:\mathbb{N}\to\{\tilde{f}(P), \tilde{f}(P')\}$ that simply look at the size of the sample $\mathcal{C}_{\tilde{f}}(\cdot)$, then we can formalise this statement in the following way.

\begin{proposition}\label{advantage}
Suppose $\tilde{f}:\datauni^*\to\mathbb{N}$ is such that for all $\eps$-DP mechanisms $\mathcal{M}$, we have that $\mathcal{M}_{\mathcal{C}_{\tilde{f}}}$ is ${\eps}'$-DP. Then for all neighboring datasets $P, P'$, we have
\[\max \adv(H; \tilde{f}(P),\tilde{f}(P'))\le e^{\eps'} - 1,\]
where the optimisation is over all hypothesis tests $H$ such that for all $x\in\mathbb{N}$, and $b\in\{0,1\}$, $e^{-\eps}\Pr(H(x)=b)\le \Pr(H(x+1)=b)\le e^{\eps}\Pr(H(x)=b)$.
\end{proposition}

The proof of this statement appears in Appendix~\ref{sec:data-dependent}. This result helps us build intuition for what type of allocation functions could possibly amplify privacy.
If $\tilde{f}$ results in privacy amplification then for any pair of neighbouring populations $P$ and $P'$, the distributions $\tilde{f}(P)$ and $\tilde{f}(P')$ must be close enough that they cannot be distinguished between by any hypothesis test $H$ such that $\log H$ is $\eps$-Lipschitz.
From this perspective the result in Section~\ref{obviousdegradation} follows from the fact that if $\tilde{f}$ is deterministic with high sensitivity then we can define an appropriate hypothesis test with large advantage (where the test statistic is $\mathcal{M}^{\text{\rm count}}$). This is a useful perspective to keep in mind throughout the remainder of the paper.

One consequence of this perspective is a lower bound on how well we can emulate a desired deterministic allocation function $f$ while controlling or amplifying privacy. Suppose that absent privacy concerns, an analyst has determined that they want to use a function $f$ to determine the sample size. However, to avoid privacy degradation they replace $f$ with a randomised function $\tilde{f}$. How close can $\tilde{f}$ get to $f$ while maintaining or amplifying the original privacy level? We can obtain a lower bound on expected closeness of $f(P)$ and $\tilde{f}(P)$ by relating it to the well studied problem of estimation lower bounds in differential privacy.

\begin{proposition}\label{var-LB}
Let $f:\datauni^*\to\mathbb{R}$ and $\eps, \eps' > 0$. Suppose $\tilde{f}:\datauni^*\to\mathbb{N}$ is a randomised function such that for all $\eps$-deletion DP mechanisms $\mathcal{M}$, it holds that $\mathcal{M}_{\mathcal{C}_{\tilde{f}}}$ is $\eps'$-replacement DP.
Let $\alpha\ge0$ be a lower bound on the $\eps'$-DP estimation rate of estimation $f$; i.e. for every $\eps'$-deletion DP mechanism $\mathcal{A}$, there exists a dataset $P$ such that $\mathbb{E}[|\mathcal{A}(P) - f(P)|^2] \ge \alpha$, then there exists a dataset $P$ such that
\[\mathbb{E}[|\tilde{f}(P) - f(P)|^2] \ge \alpha - \left(\frac{1}{\eps}\right)^2.\]
\end{proposition}
The problem of lower bounding differentially private function estimation is well-studied~\cite{Vadhan17, Asi:2020} in the privacy literature. The lower bounds essentially arise from the fact that $\mathcal{A}(P)$ and $\mathcal{A}(P')$ must be similar distributions for neighbouring databases, even if $f(P)$ and $f(P')$ are far apart. Since we know from Proposition~\ref{advantage} that $\tilde{f}(P)$ and $\tilde{f}(P')$ must also be close, we obtain the related lower bound. The slackness of $(1/\eps)^2$ is a result of the fact that while $\mathcal{A}(P)$ and $\mathcal{A}(P')$ must be indistinguishable with respect to \emph{any} hypothesis test, $\tilde{f}(P)$ and $\tilde{f}(P')$ need only be indistinguishable with respect to any $\eps$-DP hypothesis test.

\subsection{Privacy amplification from randomised rounding}\label{randomisedrounding}

Many functions used to determine data-dependent sampling rates have high sensitivity, but at least one common allocation method has low sensitivity: proportional allocation. In proportional allocation, a constant, data-independent fraction of the population is sampled independently from each stratum. This method is similar to simple random sampling, but a small amount of data dependence is introduced by the fact that the number of sampled records in each stratum must be an integer. In this section, we will show that while na\"ive implementations of proportional allocation can result in privacy degradation, a minor change in the allocation function results in privacy amplification comparable to that afforded by simple random sampling.

Let $r\in[0,1]$ and $f(P) = r|P|$ for some constant $r \in (0, 1)$. Since the output space of $f$ is not $\mathbb{N}$, it is typically, in practice, replaced with the deterministic function $\tilde{f}_{\text{det}, r}(P) = \operatorname{round}(r|P|)$, where  $\operatorname{round}( \cdot)$ rounds its input to the nearest integer.
Unfortunately, deterministic rounding can be problematic for privacy. We can see this through a simple example: suppose $P$ and $P'$ are neighbouring populations such that $|P| = 14, |P'| = 15$, and $r = 1/10$. Then, deterministic rounding always results in one case being sampled from $P$ and two cases being sampled from $P'$. As discussed in Section~\ref{obviousdegradation}, such a data-dependent deterministic function can never result in privacy amplification.

We propose a simple and practical change to the rounding process that \emph{does} guarantee roughly the expected level of privacy amplification. We replace deterministic rounding with a randomised rounding function $\tilde{f}_{\text{rand}, r}$. 
That is, let $p = r |P| - \lfloor r|P| \rfloor$ so $ \tilde{f}_{\text{rand}, r}(P) = \lceil r|P| \rceil$ with probability $p$, and $\tilde{f}_{\text{rand}, r}(P) = \lfloor r|P| \rfloor$ with probability $1 - p$.
The following proposition shows that, up to a constant factor, randomised rounding recovers the expected factor of $r$ in privacy amplification.

\begin{theorem}[Privacy Amplification from Randomised Rounding]
\label{onestratumrr} Let $r \in (0, 1)$. 
Then for every $\eps$-deletion DP mechanism $\mathcal{M}$, the mechanism $\mathcal{M}_{ \mathcal{C}_{\tilde{f}_{\text{rand}, r}}}$ is $\eps'$-deletion DP when restricted to datasets of size at least $1/r$, where \[\eps'=\log\left(1+2r(e^{2\eps}-1)\right)+\log(1+r(e^{2\eps}-1)) \approx 6r\eps.\]
\end{theorem}

The proof of this statement appears in Appendix~\ref{sec:data-dependent}. The approximation at the end of the proposition follows from applying eq. \eqref{approximation-one} and eq. \eqref{approximation-two}, which give that $\log(1+2r(\exp(2\eps) - 1)) \approx 2r \cdot 2\eps$ and $\log(1+r(\exp(2\eps) - 1)) \approx r \cdot 2\eps$. The constant $6$ can perhaps be optimized through a more careful analysis. Randomised rounding is a practical modification since it does not change the size of the sample very much; if traditional proportional allocation would assign $m$ samples, then the modified algorithm allocates at most $m+1$.

\section{Cluster sampling}\label{clustersampling}

In cluster sampling, the population is partitioned into disjoint subsets, called clusters. A subset of the clusters is sampled and data subjects are selected from within the chosen clusters. 
If the sampling scheme uses a single stage design, all data subjects contained in the selected clusters will be included in the sample. Otherwise, a random sample of data subjects might be selected from each of the selected clusters (multi-stage design).
Cluster sampling produces accurate results 
when the clusters are mutually homogeneous; that is, when the distributions within each cluster are similar to the distribution over the entire population. 

In the survey context, cluster sampling is often performed due to time or budgetary constraints which make sampling many units from a few clusters cheaper and/or faster than sampling a few units from each cluster. A typical example is when clusters are chosen to be geographic regions. Sampling a few geographic clusters and interviewing everybody in those clusters saves traveling costs compared to interviewing the same number of people based on a simple random sample from the population. 
In algorithm design, cluster sampling is often performed to improve the performance and accuracy of classifiers. In this setting, sampling often involves a two-step approach where the data is first clustered, using some clustering classifier, and then a subset of the clusters is selected. Forms of cluster samplings have been applied in several learning areas, for example in federated learning~\cite{FraboniVKL21} and active learning~\cite{NguyenS04}. 

\subsection{Privacy implications of single-stage cluster sampling with simple random sampling}
\label{sec:simple-clustering}
We focus here on a simple cluster sampling design that is commonly used in survey sampling and which na\"ively appears to be a good candidate for privacy amplification: simple random sampling without replacement of clusters. Suppose the dataset $P$ is divided into $k$ disjoint clusters,
\[P=C_1\sqcup\cdots\sqcup C_k\]
and the sampling mechanism $\mathcal{C}_{\clusterrate}:\datauni^*\to\datauni^*$ chooses a random subset $I\subset [k]$ of size $\clusterrate<k$, then maps $P$ to $\sqcup_{i\in I} C_i$. 

Since simple random sampling at the individual level provides good privacy amplification, one might  expect the same to happen when the clusters are sampled in a similar way.  In fact, this is true when the size of each cluster is small. 
However, if the clusters are large, this sampling design achieves less amplification than might be expected.  This is characterized by the 
following theorem showing a lower bound in this setting. 

\begin{theorem}[Lower Bound on Privacy Amplification for Cluster Sampling]\label{clusterLB}
For any sequence $n_i>0$ and privacy parameter $\eps>0$, there exist neighboring populations $P=C_1\sqcup\cdots\sqcup C_i\sqcup\cdots\sqcup C_k$ and $P'=C_1\sqcup\cdots\sqcup C_i'\sqcup\cdots\sqcup C_k$ (with $|C_i|=n_i$ and $C_i'=C_i\cup\{x\}$ for some $x\in\datauni$) and an $\eps$-deletion DP mechanism $\mathcal{M}$ such that if $\mathcal{M}_{\mathcal{C}_{\clusterrate}}(P)$ and $\mathcal{M}_{\mathcal{C}_{\clusterrate}}(P')$ are $\eps'$-indistinguishable then 
\[\eps'\ge\ln\left(1+\frac{\frac{\clusterrate}{k}}{\left(\frac{\clusterrate}{k}+\left(1-\frac{\clusterrate}{k}\right)e^{-(n_i+n_{\min})\eps}\right)}(e^{\eps}-1)\right),\] where $n_i=|C_i|$ and $n_{\min}=\min_{j\in\{1,\cdots,i-1\}\cup\{i+1,\cdots,k\}}n_j$.
\end{theorem}

We can compare the expression in the theorem above with the upper bound we have for simple random sampling without replacement (cf. Theorem 14 from \cite{Balle:2018}):
\begin{equation}\label{simpleamplfication}
\eps'=\ln\left(1+\frac{m}{n}(e^{\eps}-1)\right),
\end{equation}
where $m$ samples are drawn from a population of size $n$. 
Let us consider the case in which all the clusters are small. In this case, the quantity $n_i+n_{\min}$ will also be small, and if $\eps<1$, we can still 
expect some privacy amplification. However, as the clusters grow in size, the quantity $n_i+n_{\min}$ will also increase, and the lower bound converges very quickly to $\eps$, giving essentially no amplification. 

Next, we present a corresponding upper bound.
 \begin{theorem}[Upper Bound on Privacy Amplification for Cluster Sampling]\label{clusterUB}

For any sequence $n_i>0$, privacy parameter $\eps>0$,  $\eps$-deletion DP mechanism $\mathcal{M}:\datauni^*\to\mathcal{O}$, and pair of neighboring populations $P$ and $P'$ such that 
$P=C_1\sqcup\cdots\sqcup C_i\sqcup\cdots\sqcup C_k$ and $P'=C_1\sqcup\cdots\sqcup C_i'\sqcup\cdots\sqcup C_k$ (with $|C_i|=n_i$ and $C_i'=C_i\cup\{x\}$ for some $x\in\datauni$), 
the random variables $\mathcal{M}_{\mathcal{C}_{\clusterrate}}(P)$ and $\mathcal{M}_{\mathcal{C}_{\clusterrate}}(P')$ are $\eps'$-indistinguishable where \[\eps'\le\ln\left(1+\frac{\frac{\clusterrate}{k}}{\left(\frac{\clusterrate}{k}+\left(1-\frac{\clusterrate}{k}\right)e^{-(n_i+n_{\max})\eps}\right)}(e^{\eps}-1)\right),\] and  $n_{\max}=\max_{j\in\{1,\cdots,i-1\}\cup\{i+1,\cdots,k\}}n_j$,
\end{theorem}

Similar to the lower bound, the upper bound will quickly approach $\eps$ if the quantity $n_i+n_{\max}$ is large. 
If each cluster contains a single data point, 
the two bounds are close. This is not surprising since in this case the type of cluster sampling we considered is just simple random sampling without replacement. Note that while $\clusterrate/k$ is the fraction of clusters included in the final sample and $m/n$ is the fraction of data points, these are approximately the same when the clusters are small.
  If all the clusters are the same size, then $n_{\max}=n_{\min}$ and the upper and lower bounds in Theorem~\ref{clusterLB} and~\ref{clusterUB} match. Once again it is worth comparing the expression in the theorem above eqn~\eqref{simpleamplfication}. The discrepancy in the case when $n_1=1$ for all $i$ is due to moving from $\mathcal{M}$ satisfying $\eps$-DP in the replacement model to $\eps$-DP in the deletion model.
 The proofs of these results are contained in Appendix~\ref{app:cluster}. The following section gives some further intuition for these results.
 
\subsection{Discussion and hypothesis testing}\label{hypothesistestingandclustersampling}
Privacy amplification by subsampling is often referred to as \emph{secrecy of the sample} due to the intuition that the additional privacy arises from the fact that there is uncertainty regarding which user's data is in the sample.
The key intuition then for Theorem~\ref{clusterLB} is that the larger the clusters are, the easier it is for a differentially private algorithm $\mathcal{M}$ to reverse engineer which clusters were sampled, breaking secrecy of the sample.
Intuitively, if the clusters are different enough that a private algorithm can guess which clusters were chosen as part of the sample, then any amplification due to \emph{secrecy of the sample} is negligible. We can formalize this intuition using once again the lens of hypothesis testing. Note that the framing in this section differs slightly from the framing in Section~\ref{datadependentsampling}, although the underlying idea in both settings is that if a particular hypothesis test is effective, then this implies a lower bound on the privacy parameter.
In addition, note that privacy is also conserved in this setting, as $\mathcal{M}_{\mathcal{C}_{\clusterrate}}$ is at least as private as $\mathcal{M}$. The question is: when is $\mathcal{M}_{\mathcal{C}_{\clusterrate}}$ \emph{more} private than $\mathcal{M}$?

\begin{theorem}\label{clusterHT}
Let $\eps>0$, $\clusterrate\in[0,k]$, $\mathcal{M}:\datauni^*\to\outputspace$ be $\eps$-DP and the sampling mechanism $\mathcal{C}_{\clusterrate}$ be as defined in Section~\ref{sec:simple-clustering}. Suppose there exists a hypothesis test $\mathcal{H}:\outputspace\to\{0,1\}$ such that \[\Pr(\mathcal{H}(\mathcal{M}_{\mathcal{C}_{\clusterrate}}(P))=0\;|\; C_i\in \mathcal{C}_{\clusterrate}(P))\ge e^{\eps'}\Pr(\mathcal{H}(\mathcal{M}_{\mathcal{C}_{\clusterrate}}(P))=0\;|\; C_i\notin \mathcal{C}_{\clusterrate}(P)).\]
Then there exists an event $E$ in the output space of $\mathcal{M}$ such that for any neighboring population $P'$ that differs from $P$ in $C_i$, if \[\eps''=\log\frac{\Pr( \mathcal{M}_{\mathcal{C}_{\clusterrate}}(P) \in E \vert C_i \in \mathcal{C}_{\clusterrate}(P))}{\Pr( \mathcal{M}_{\mathcal{C}_{\clusterrate}}(P') \in E \vert C_i \in \mathcal{C}_{\clusterrate}(P'))}\in [0,\eps],\] and $\mathcal{M}_{\mathcal{C}_{\clusterrate}}(P)$ and $\mathcal{M}_{\mathcal{C}_{\clusterrate}}(P')$ are $\tilde{\eps}$-indistinguishable, then \[\tilde{\eps} \ge \log\left(1+(e^{\eps''}-1)\frac{\clusterrate/k}{\clusterrate/k+e^{-\eps'}(1-\clusterrate/k)}\right).\] 
\end{theorem}

The key take-away of this theorem is that for any $\eps$-DP mechanism $\mathcal{M}$, if there exists a hypothesis test that, when given the output of $\mathcal{M}_{\mathcal{C}_{\clusterrate}}(P)$, can confidently decide whether cluster $C_i$ was chosen as part of the final sample, then the privacy guarantee of $\mathcal{M}_{\mathcal{C}_{\clusterrate}}$ is no better than the privacy guarantee would be if we knew for certain that $C_i$ was chosen as part of the sample. That is, in this setting, we gain no additional privacy as a result of secrecy of the sample. The parameter $\eps'$ controls how well the hypothesis test can determine whether $C_i\in\mathcal{C}_{\clusterrate}$. As $\eps'$ increases, $\tilde{\eps}$ approaches $\eps''$, the privacy parameter if $C_i$ is known to be part of the sample, so privacy amplification is negligible. We will generalise this intuition to a broader class of sampling designs in Theorem~\ref{PPSHT}.

This view is consistent with Theorem~\ref{clusterLB}. Consider a population where only data points in cluster $i$ have a particular property and let $\mathcal{M}$ be an $\eps$-DP mechanism that attempts to count how many data points with the property are in the final sample. If cluster $i$ is large, then it is easy to determine from the output of the mechanism whether $C_i$ is in the final sample.
This example required cluster $i$ to be distinguishable from the remaining clusters using a private algorithm. While examples as extreme as the one above may be uncommon in practice, clusters being different enough for a private algorithm to distinguish between them is not an unrealistic assumption.

In Section~\ref{sec:simple-clustering}, we analysed a single stage design. All subjects contained in the selected clusters were included in the sample. In practice, multi-stage designs are common, where a random sample of subjects are selected from within each chosen cluster. If the sampling within each cluster is sufficiently simple then the privacy amplification from this stage can be immediately incorporated into the upper bound in Theorem~\ref{clusterUB}. For example, if each subject within the chosen clusters is sampled with probability $r$ and $\mathcal{M}$ is $\eps$-DP, i.e., we perform Poisson sampling with probability $r$,
then we immediately obtain an upper bound that is approximately $r\eps$.
However, if the sampling within clusters is more complex, then further analysis is required. One can also imagine more complicated schemes for selecting the chosen clusters. We study sampling the clusters with probability proportional to their size (PPS), which is another popular cluster sampling design, in Section~\ref{sec:PPS_cluster}. 

\section{Analysing Other Data-dependent Sampling Schemes}\label{othersamplingschemes}

In this section, we highlight the limitations of privacy amplification for other data-dependent sampling schemes. Our results follow from a generalization of Theorem~\ref{clusterHT} and provide a lower bound on privacy amplification for general sampling schemes.
This allows us to draw out key properties 
of sampling designs that hinder privacy amplification, such as the probabilities of selection being sensitive to changes in the dataset.
 We apply our result to three common schemes: Probability Proportional to Size (PPS) sampling on the element level, PPS sampling of clusters, and systematic sampling. These sampling schemes are widely used in practice to improve efficiency, that is, to reduce the variability of the obtained estimates. We discuss lower bounds on privacy amplification for each of these three schemes.

The main general result of this section is Proposition~\ref{PPSHT}, below, which generalizes the intuition of Theorem~\ref{clusterHT} and provides a lower bound on privacy amplification of general sampling schemes. We will discuss the implications of the result after stating it.

\begin{proposition}\label{PPSHT}
Let $\eps>0$,  $\mathcal{M}:\datauni^*\to\outputspace$ be any mechanism and $\mathcal{C}:\datauni^n\to\datauni^*$ be any sampling mechanism. Let $P=\{x_1,\cdots,x_n\}$ and $P'=\{x_1',\cdots,x_n'\}$ be two neighboring datasets, such that $x_i=x_i'$ for all $i\in[2:n]$. For all $i\in[n]$, let 
\begin{itemize}
    \item $a_i = \Pr(x_i \in \mathcal{C}(P))$ and $a'_i = \Pr(x_i' \in \mathcal{C}(P'))$ be the probabilities of inclusion of element $i$,
    \item $p_i(E)=\Pr(\mathcal{M}_{\mathcal{C}}(P)\in E\;|\; x_i\in \mathcal{C}(P))$ 
        and $p_i'(E)=\Pr(\mathcal{M}_{\mathcal{C}}(P')\in E\;|\; x_i'\in \mathcal{C}(P'))$ be the distribution on outputs conditioned on element $i$ being chosen,
    \item $q_i(E)=\Pr(\mathcal{M}_{\mathcal{C}}(P)\in E\;|\; x_i\notin \mathcal{C}(P))$ and $q_i'(E)=\Pr(\mathcal{M}_{\mathcal{C}}(P')\in E\;|\; x_i'\notin \mathcal{C}(P'))$ be the distribution on outputs conditioned on element $i$ not being chosen.
    \end{itemize}
and  $\mathcal{M}_{\mathcal{C}}(P)$ and $\mathcal{M}_{\mathcal{C}}(P')$ are $\tilde{\eps}$-indistinguishable, then
\begin{align}\label{privLB}
    \tilde{\eps} \geq \max_{E, i \in [n]} \ln \left( \frac{p_i(E)}{p_i'(E)} \cdot \frac{a_i}{a_i'} \cdot \frac{1}{1+\frac{q_i'(E)(1-a_i')}{p_i'(E)a_i'}} \right)
\end{align}
\end{proposition}

Proposition~\ref{PPSHT} generalises the intuition of Theorem~\ref{clusterHT} to a broader class of sampling schemes, and to any mechanism. The mechanism $\mathcal{M}$ is not required to be $\eps$-DP but this theorem is most interesting if we consider $\mathcal{M}$ to be $\eps$-DP. In this case, we expect $p_i(E)/p_i'(E)$ to be around $e^{\eps}$
, although it can be lower if element $i$ has little impact on the outcome, or higher depending on the structure of the sampling scheme. We can think of this ratio as the privacy loss guarantee if we remove secrecy of the sample for element $i$, so this element acts as something like a upper bound for the privacy loss. The proposition lower bounds how far $\tilde{\eps}$ can be from this baseline privacy loss.

In particular, the result highlights three ways that an algorithm that composes a sampling mechanism and a differentially private mechanism can fail to achieve privacy amplification by sampling:

\begin{itemize}
\item Probability of inclusion close to 1. When the ratio $(1-a_i')/a_i'$ is small, then the lower bound on $\tilde{\eps}$ is large.
Notice that if $a_i=a_i'=1$ then, as expected, there is no amplification (i.e.  $\tilde{\eps}=\max_{E,i\in[n]} \ln (p_i(E)/p'_i(E))$). 
\item Probability of inclusion is data dependent. If $a_i/a_i'$ is large, then privacy degradation can actually occur, i.e. $\tilde{\eps}$ can exceed $p_i(E)/p'_i(E)$. While this ratio plays a role in controlling the privacy loss, it alone does not determine $\tilde{\eps}$, and privacy amplification can occur even if the ratio is large.
\item The final ratio $q'_i(E)/p_i'(E)$ is related to how much information the output event $E$ contains about whether or not element $i$ was in the data set. If the event is much more likely if element $i$ is in the data set (the ratio is small), then this can prevent privacy amplification. Note that this ratio can be larger than $\eps$ due to correlations in the sampling. For example, in cluster sampling, if a subject $i$ is included in the sample, all of the elements in that cluster are included, which is easily detectable. 
\end{itemize}

\begin{proof}[Proof of Proposition~\ref{PPSHT}] 
Let $D = \mathcal{C}(P)$ and $D' = \mathcal{C}(P')$. Then,
\begin{align}
    \nonumber e^{\tilde{\eps}} \ge \frac{\Pr( \mathcal{M}_{\mathcal{C}}(P) \in E) }{\Pr (\mathcal{M}_{\mathcal{C}}(P') \in E )} 
    &= \frac{\Pr( \mathcal{M}(D) \in E) }{\Pr (\mathcal{M}(D') \in E )} \\
    &= \frac{p_i(E) \cdot a_i + q_i(E) \cdot (1-a_i)}{p'_i(E) \cdot a_i' + q'_i(E) \cdot (1-a'_i)} \label{numericalLB} \\
    \nonumber &\ge \frac{p_i(E)}{p_i'(E)} \cdot \frac{a_i}{a_i'} \cdot \frac{1}{1+\frac{q'(E)(1-a_i')}{p'(E)a_i'}}.
\end{align}

\end{proof}

Proposition~\ref{PPSHT} generalises the intuition of Theorem~\ref{clusterHT} but is not a direct generalisation of the theorem itself. In particular, Theorem~\ref{clusterHT} finds a particular mechanism $\mathcal{M}$ and event $E$ for which the privacy guarantee of $\mathcal{M}_{\mathcal{C}}$ is lower bounded.
We are able to obtain a stronger lower bound for cluster sampling by taking advantage of the structure of the sampling design. 

For the remainder of this section we will consider the implications of Proposition~\ref{PPSHT} for several important sampling schemes: probability proportional to size sampling on the element level, probability proportional to size sampling on the cluster level, and systematic sampling.

\subsection{Probability Proportional to Size (PPS) Sampling}\label{PPS}

Probability Proportional to Size (PPS) sampling is a sampling method where 
each element $x_i, i \in [n]$ has a probability of inclusion in the final sample which is proportional to a quantity $r_i$ associated with subject $i$. The quantities $r_1, \ldots, r_n$ may correspond to an auxiliary variable, such as the size of businesses in an establishment survey. PPS sampling can improve the accuracy of the final population estimate over simple random sampling by re-weighting data points to ensure that more influential data points have a higher probability of inclusion. As implied by the name, the auxiliary variable $r_i$ is often related to the size or influence of a data point. However, everything we say in this section will hold for any data dependent sampling scheme. 

Applying Proposition~\ref{PPSHT}, it is easy to see that the privacy amplification of PPS sampling is capped by the highest probability of inclusion possible for any data subject. To see this, let $P=\{x_1,\cdots,x_n\}$ be a data set of size $n$ and $i\in[n]$. Let $P'$ be any neighboring data set that differs from $P$ on the data of individual $i$. Let $\mathcal{M}(S)$ be the algorithm that samples from $\texttt{Bernoulli}(e^{\eps}/(e^{\eps}+1))$ if $x_i$ is in $S$, and samples from $\texttt{Bernoulli}(1/(e^{\eps}+1))$ otherwise. Then, using the notation from Proposition~\ref{PPSHT}, and letting $E=\{1\}$, we have $p_i(E)=e^{\eps}/(e^{\eps}+1)$, $p'_i(E)$
$=q_i(E)=q_i'(E)=1/(e^{\eps}+1)$. Thus, $\tilde{\eps}\ge \max_{i\in[n]} \ln(e^{\eps}a_i)$. By using eqn~\eqref{numericalLB} directly, we can obtain the stronger statement that \[\tilde{\eps}\ge\max_{i\in[n]} \ln(1+a_i(e^{\eps}-1)).\]
There are several methods used to achieve, or nearly achieve PPS sampling. This lower bound on privacy amplification by sampling for PPS is agnostic to the particular sampling method used. How close the privacy loss can be upper bounded is dependent on the particular sampling method.

The existence of a large amount of deviation in the $r_i$'s can imply the existence of a large $a_i$, thus preventing strong privacy amplification. 
Privacy loss can be controlled by limiting the variation in the $r_i$. For example, if all the $r_i$ are equal then PPS sampling is exactly simple random sampling without replacement, and so achieves privacy amplification, but the accuracy gains from PPS sampling will be lost. 

\subsection{PPS Cluster Sampling}\label{sec:PPS_cluster}

PPS sampling is also commonly used for cluster sampling when the clusters are of varying sizes. When combined with simple random sampling with appropriate sampling rates within the chosen clusters, sampling each cluster with probability proportional to its size ensures that each individual in the population has an equal probability of being included in the final sample, while maintaining the practical benefits of cluster sampling.

Let us first consider cluster sampling as a single step design, without simple random sampling within clusters. As in Section~\ref{clustersampling}, suppose the data set $P$ is divided into $k$ clusters,
$P=C_1\sqcup\cdots\sqcup C_k.$
The sampling mechanism $\mathcal{C}_{\rm PPS}:\datauni^*\to\datauni^*$ chooses cluster $i$ with probability proportional to the size of the cluster, $|C_i|$, and returns  cluster $i$ as the sample. 
We can again use Theorem~\ref{PPSHT} to analyse this sampling scheme. 
As in standard cluster sampling, if the clusters are large then secrecy of the sample is broken ($\eps'$ is large) and no privacy amplification is achieved. Recall the analysis we did in Section~\ref{hypothesistestingandclustersampling} of standard cluster sampling, where we found a specific $\eps$-DP mechanism $\mathcal{M}$ then lower bounded the privacy guarantee of $\mathcal{M}_{\mathcal{C}}$. We can use the same mechanism and pair of datasets, combined with Theorem~\ref{PPSHT} to give an example of a mechanism that amplified poorly when combined with PPS cluster sampling.
As in standard cluster sampling, if the clusters are large then secrecy of the sample is broken ($\eps'$ is large) and no privacy amplification is achieved. The existence of a single large cluster also means that this cluster is more likely to be chosen, further pushing us towards the no amplification regime ($a_i$ and $a_i'$ are close to 1). 
Privacy loss can be controlled by limiting the size of clusters, although this solution has practical implications as PPS sampling is commonly used in situations in which the sampler has little control over the size of the clusters.

\subsection{Systematic Sampling}

Systematic sampling is a sampling method commonly used as a practical replacement for sampling techniques that are difficult to implement in practice, e.g. simple random sampling without replacement or PPS sampling. When used in place of simple random sampling, \emph{systematic sampling} imposes an ordering on the population, randomly selects an individual among the first $k$ individuals, where $k= |P|/m$\footnote{If this ratio is not an integer than various techniques can be employed to maintain a sampling design with equal probabilities of selection. For this work, we will assume that $k$ is an integer.}, then selects every $k$th individual thereafter. The ordering on the population can be generated in a number of ways. In particular, the ordering may include randomness (e.g. every $k$th consumer entering a store) or not (e.g. by establishment size). This technique can be generalised to any desired sampling distribution while (nearly) maintaining the marginal probability of each individual being chosen. To see this, imagine expanding the population so that individuals with higher probability of being chosen are ``duplicated", then perform the systematic sampling described above to this expanded population. In addition to being a practically simpler sampling technique, systematic sampling has the additional benefit of implicitly imposing some stratification, if the population is sorted by some known attributes. We will focus on systematic sampling for simple random sampling in this section since the main ideas are already present in this example. The privacy implications of systematic sampling are very different to that of simple random sampling. Namely, the privacy amplification achieved by systematic sampling is negligible for standard parameter settings. 

The amount of uncertainty in the ordering of the population affects the privacy implications of systematic sampling.
If the ordering is uniformly random, that is the population is uniformly randomly ordered before sampling and the ordering is kept secret, then systematic sampling results in exactly the same sampling distribution as simple random sampling. Hence systematic sampling with random ordering achieves privacy amplification as in Theorem~\ref{priorwork}. However, in the other extreme, where there is no randomness in the ordering, we can view systematic sampling as almost an instance of cluster sampling.
Suppose that there exists a total ordering on all possible data subjects. That is, for any possible individuals $i$ and $i'$, either $i<i'$, or $i'<i$. For example, imagining ordering data subjects alphabetically. Then we can view systematic sampling as first creating $k$ ``clusters", where cluster $j$ contains individual $j, k+j, 2k+j, \cdots$, then sampling a cluster uniformly at random. From this perspective, we can see that systematic sampling suffers from the same phenomenon as cluster sampling. If the desired sample size is large enough to detect which cluster was chosen with a differentially private algorithm, then there is no amplification (assuming the ordering in the population is known to the attacker).

\section{Stratified sampling} \label{sec:stratified}

Finally, we turn our attention to another common sampling design: stratified sampling.
In stratified sampling, the data is partitioned into disjoint subsets, called strata. A subset of data points is then sampled from each stratum to ensure the final sample contains data points from every stratum. Stratified sampling is common in survey sampling where it is used to improve accuracy and to ensure sufficient representation of sub-populations of interest. 
A classic use case of stratified sampling is business surveys, where businesses are typically stratified by industry and number of employees, or by similar measures of establishment size. Stratification by establishment size results in substantial gains in accuracy compared to simple random sampling, while stratification by industry ensures that reliable estimates can be obtained at the industry level. Stratified sampling has several other applications; for example it is used in algorithm design to improve performance~\cite{AlafateF19,KoolvHW20}, 
in private query design and optimization to improve accuracy~\cite{BaterP0WR20}, and to improve search and optimizations~\cite{LelisSAZFH16}. 

We focus here on \emph{one-stage stratified sampling} using simple random sampling without replacement within each stratum to select samples. We also assume that the stratum boundaries have been fixed in advance. Given a target sample size $m$, the only design choice in this model is the \emph{allocation function}, which determines how many samples to take from each stratum. Different allocation functions are used in practice. Which method is selected depends on the goals to be achieved (for example, ensuring constant sampling rates across strata or minimizing the variance for a statistic of interest).

Before we describe allocation functions in detail, let us establish some notation for stratified sampling. Suppose there are $k$ strata in the population, and that each data point is a pair $(s, x)$ where $s~\in~[k]$ denotes which stratum the data subject belongs to, and $x\in\mathcal{U}$ denotes their data. 
Let $\bm{f} = (\bm{f}_1, \dots, \bm{f}_k) : ([k] \times \datauni)^* \to \mathbb{N}^k$ denote the allocation rule, so $\bm{f}_i(P)$ samples are drawn uniformly at random without replacement from the $i$th stratum, $P_i=\{(s,x)\in P\;|\; s=i\}$. The final sample $S$ is the union of the samples from all the strata. 

An important  feature of stratified sampling is that the sampling rates can vary between the strata. This means that data subjects in strata with low sampling rates may expect a higher level of privacy than data subjects in strata with high sampling rates. This leads us to define a variant of differential privacy that allows the privacy guarantee to vary between the strata. This generalisation of differential privacy is tailored to stratified datasets and allows us to state more refined privacy guarantees than the standard definition is capable of.

\begin{definition}
Let $k\in\mathbb{N}$ and suppose there are $k$ strata. A mechanism $\mathcal{A}$ satisfies $(\eps_1, \cdots, \eps_k)$-stratified replacement differential privacy if for all datasets $P$, data points $(s, x)$ and $(s',x')$, $\mathcal{A}(P\cup\{(s,x)\})$ and $\mathcal{A}(P\cup\{(s',x')\})$ are $\max\{\eps_s, \eps_{s'}\}$-indistinguishable. The mechanism $\mathcal{A}$ satisfies $(\eps_1, \cdots, \eps_k)$-stratified deletion differential privacy if for all datasets $P$, data points $(s, x)$, $\mathcal{A}(P)$ and $\mathcal{A}(P\cup\{(s,x)\})$ are $\eps_s$-indistinguishable.
\end{definition}

This definition is an adaptation of
personalized differential privacy \cite{Jorgensen:2015, Ebadi:2015, Alaggan:2016}. Note that it protects not only the \emph{value} of an individual's data point, but also which stratum they belong to. 

\subsection{Optimal allocation with privacy constraints}
In this section, we will discuss how to think about choosing an allocation function when privacy is a concern. A common goal when choosing an allocation $\bm{f}$ is to minimise the variance of a particular statistic. That is, suppose that $\mathcal{C}_{\bm{f}}$ represents one-stage stratified sampling with allocation function $\bm{f}$. Then, given a population $P$ and desired sample size $m$, the \emph{optimal allocation function} $\bm{f}^*(P)$ with respect to a statistic $\theta$ is defined as
\begin{equation}\label{neymanallocdefn}
\bm{f}^*(P) = \arg\min_{\bm{f}} \text{var}(\theta_{\mathcal{C}_{\bm{f}}}(P)),\end{equation}
where the randomness may come from both the allocation function and the sampling itself, and the minimum is over all allocation functions such that $\|\bm{f}(P)\|_1\le m$ for all $P$. \footnote{We note that the notion of optimal allocations implicitly assumes that the historic or auxiliary data, $H$, used to inform the sampling design and the population data $P$ are the same, or at least similar enough that $\bm{f}^*(H)$ is a good proxy for $\bm{f}^*(P)$. This provides further justification for the assumption that $H=P$ in our statements.}

A natural question then is: what is the optimal allocation when one wants to compute the statistic of interest differentially privately?  
This is a simple yet subtle question. Our results in the previous sections indicate that the landscapes of optimal allocations in the non-private and private settings may be very different. This is a result of the fact that allocation functions that do not amplify well typically need to add more noise to achieve privacy (see discussion in Section~\ref{howtouse}). The additional noise needed to achieve privacy may overwhelm any gains in accuracy for the non-private statistic. Additionally, it is not immediately obvious how to define the optimal allocation in the private setting.

In this section, we formulate the notion of an optimal allocation under privacy constraints. Our goal is to initiate the study of alternative allocation functions that may prove useful when privacy is a concern. A full investigation of this question is outside the scope of this paper, but we provide some intuition for why this may be an interesting and important question for future work. 

Given a statistic $\theta$, we wish to define the optimal allocation for estimating $\theta$ privately. Let $\tilde{\theta}^{\lambda}$ be an $\lambda$-DP algorithm for estimating $\theta$, so $\tilde{\theta}^{\lambda}(P)$ is an approximation of $\theta(P)$. The smaller $\lambda$ is, the noisier $\tilde{\theta}^{\lambda}$ is.
The scale of $\lambda$ needed to ensure that ${\tilde{\theta}^{\lambda}}_{\mathcal{C}_{\bm{f}}}$ is $\eps$-DP depends on the allocation function $\bm{f}$.
Allocation functions that are very sensitive to changes in the input dataset will require more noise (smaller $\lambda$) to mask changes in the allocation. 
For any allocation $\bm{f}$, we will define the optimal parameter $\lambda$ as that which minimises the maximum variance of ${\tilde{\theta}^{\lambda}}_{\mathcal{C}_{\bm{f}}}$ over all datasets $P$, while maintaining privacy:
 \begin{align}\lambda_{\bm{f}} = &\argmin{\lambda>0}\sup_{P}\frac{\text{var}({\tilde{\theta}^{\lambda}}_{\mathcal{C}_{\bm{f}}}(P))}{\text{var}(\theta_{\mathcal{C}_{\bm{f}}}(P))}\label{noiseafteramp}\\
\nonumber&\hspace{0.1in}\text{s.t. } {\tilde{\theta}^{\lambda}}_{\mathcal{C}_{\bm{f}}} \text{ is } (\eps_1, \cdots, \eps_k)\text{-stratified DP.}\end{align} Now, by definition, ${\tilde{\theta}^{\lambda_{\bm{f}}}}_{\mathcal{C}_{\bm{f}}}$ is $(\eps_1, \cdots, \eps_k)$-stratified DP for any allocation function $\bm{f}$. We minimise the multiplicative increase in variance so that the supremum is not dominated by populations $P$ for which $\text{var}(\theta_{\mathcal{C}_{\bm{f}}}(P))$ is large.
Given privacy parameters $\eps_1, \cdots, \eps_k\ge 0$, we now define the optimal allocation as the allocation function that minimises the maximum variance over all populations $P$: \begin{align}\label{privateoptimal}
\bm{f}^*_{\eps} &= \underset{\bm{f}}{\operatorname{argmin}}\sup_P\text{var}({\tilde{\theta}^{\lambda_{\bm{f}}}}_{\mathcal{C}_{\bm{f}}}(P)).
\end{align}
where the minimum again is over all allocations $\bm{f}$ such that $\|\bm{f}(P)\|_1\le m$ for all $P$, and the supremum is over all populations of interest. This optimisation function has a different form to Eqn~\ref{neymanallocdefn}, which performs the optimisation independently for each population $P$. 
This difference is necessary in the private setting as we need to ensure that the choice of allocation function $\bm{f}^*_{\eps}$ is not data dependent, since this would introduce additional privacy concerns.
We can view the optimal allocation as the optimal balancing between the variance of the non-private statistic, and the scale of the noise needed to maintain privacy. 

We believe that examining the difference between the optimal allocation in the non-private setting (Eqn~\eqref{neymanallocdefn}) and in the private setting (Eqn~\eqref{privateoptimal}) is an important question for future work. 
The main challenge is computing the parameter $\lambda_{\bm{f}}$ for every allocation $\bm{f}$. Analysing the privacy implications of $\bm{f}$ in the style of the previous sections gives us an upper bound on $\lambda_{\bm{f}}$, although this bound may be loose for specific statistics $\tilde{\theta}^{\lambda}$. So, while the previous sections developed our intuition for $\lambda_{\bm{f}}$, we believe new techniques are required to understand this parameter enough to solve Eqn~\eqref{privateoptimal}.

\subsection{Challenges with optimal allocation}

Optimal allocations are defined to perform well for a specific statistic of interest. However, in practice, a wide variety of analyses will be performed on the final sample. The chosen allocation function may be far from optimal for these other analyses. While this problem exists in the non-private setting, it becomes more acute in the private setting. An allocation function that is optimal for one statistic may result in privacy degradation (and hence low accuracy estimates) for another. 

We illustrate this challenge using \emph{Neyman allocation}, which is often employed for business surveys. Neyman allocation is the optimal allocation method for the weighted mean~\cite{Neyman1934}: \[\weightedmean(S) = \frac{1}{|P|}\sum_{i=1}^k \frac{|P_i|}{|S_i|}\sum_{x\in S_i} x,\] where $|P_i|$ is the size of stratum $i$, and $S_i=S\cap P_i$. The estimator $\weightedmean(S)$ is an unbiased estimate of the population mean for any stratified sampling design. Given a desired sample size $m$, let $\neymanalloc$ be the allocation
function corresponding to Neyman allocation. Provided each stratum is sufficiently large, $\neymanalloc(P) = (m_1, \cdots, m_k)$, where \[m_i=\frac{ |P_i| \sigma(P_i) }{ \sum_{j=1}^k |P_j| \sigma(P_j)} \cdot m,\] $\sigma^2(P_i)$ is the empirical variance in stratum $i$ and sufficiently large means that $m_i\le |P_i|$. 
Neyman allocation is deterministic and can be very sensitive to changes in the data due to its dependence on the variance within each stratum. So, while it can provide accurate results for some statistics, it provides very noisy results for other statistics of potential interest (e.g. privately computing strata sizes).

To demonstrate the sensitivity of Neyman allocation, we analyze its outputs on a simulated data set. We consider a population based on the County Business Patterns (CBP) data published by the U.S. Census Bureau~\cite{eckert2021}.\footnote{The data released by the U.S. Census Bureau is a tabulated version of
the underlying microdata from the Business Register (BR), a database of all known single and multi-establishment employer companies. We generate a simulated data set that is consistent with the tabulated release. In order to compute the sensitivity of Neyman allocation, we top-code the establishment size at 10,000.} Each row of the data set corresponds to an establishment, and the establishments are stratified by establishment size into $k=12$ strata. With a target final sample size of $m=10,000$, and using the weighted mean of the establishment size as the target statistic, the Neyman allocation for this population is $[1261, 621, 517, 1969, 833, 1947, 1058, 762, 257, 248, 306, 225]$. We can find a neighbouring population with Neyman allocation $[1259, 620, 516, 1965, 831, 1943, 1056, $ $761, 257, 247, 306, 244]$. While these allocations are not wildly different, they do differ by 19 samples in the top stratum, which might not have a large impact on the weighted mean, but could lead to more substantial changes for other statistics. As an illustrative example, we can consider the goal of privately estimating the stratum sizes in the sample. For such a goal, this allocation would lead to significant privacy degradation. 
See Appendix~\ref{app:neyman-CBP} for more details.

\subsection{Privacy amplification from proportional sampling}
\emph{Proportional sampling} is an alternative allocation function that is used to provide equitable representation of each sub-population, or stratum. Given a desired sample size $m\in[n]$, proportional sampling samples an $r=\frac{m}{n}$ fraction of the data points (rounded to an integer) from each stratum. Proportional sampling is not an optimal allocation in the non-private setting but, when implemented with randomised rounding, it has good privacy amplification.
Now that we consider stratified sampling with number of strata $k \geq 1$, we can state the following generalisation of Theorem~\ref{onestratumrr}.

\begin{theorem}[Privacy Amplification for Proportional Sampling] \label{sswor} 
Let $r\in[0,1]$, $\eps>0$, $\mathcal{M}$ be an $\eps$-DP mechanism,
and $P=S_1\sqcup\cdots\sqcup S_k$ and $P'=S_1'\sqcup\cdots\sqcup S_k'$ be stratified neighboring datasets that differ on stratum $i$. 
If for all $j\in[k]$, $r|S_j|\ge 1$ and $r|S_j'|\ge 1$, then $\mathcal{M}_{\mathcal{C}_{\bm{f}_{r,\text{\rm prop}}}}$ is $\eps'$-DP where 
\[\eps'\le\log\left(1+2r(e^{2\eps}-1)\right)+\log(1+r(e^{2\eps}-1)).\]
\end{theorem}

Note that given a private statistic $\tilde{\theta}^{ \lambda}$ as defined as above, this allows us to set $\lambda_{\bm{f}_{r,\text{prop}}}\approx \frac{\eps}{6r}$, which is considerably larger than $\eps$ for small sampling rates. Thus, while proportional sampling may not minimise the variance of any single statistic, it may be a good choice since it performs reasonably well for \emph{all} statistics.

\section{Conclusion}

In this paper, we have considered the privacy guarantees of sampling schemes, extending previous results to more complex and data-dependent sampling designs that are commonly used in practice. We find that considering these sampling schemes requires developing more nuanced analytical tools. In this work, we analyse the privacy impacts of randomized and data-dependent sampling schemes. Then, we apply our insights to a variety of sampling designs. To the best of our knowledge, this work is the first to initiate study into these designs. As such, we hope to foster further study of the interplay between sampling designs and private algorithms. In particular, the results in this paper indicate that optimal allocation functions under a differential privacy constraint may be different to optimal allocations in the non-private setting and warrants further study. We also observed that mechanisms that incorporate information from the sampling design (e.g. sample weights) pose a unique challenge for private analysis and new techniques need to be developed to study these mechanisms.

\bibliographystyle{acm}
\bibliography{biblio}
\newpage
\appendix

\onecolumn
\appendix

\section{Basic facts about indistinguishability}

\begin{definition}
Let the \emph{LCS distance} between two data sets $P$ and $P'$, denoted $\LCS(P,P')$, be the minimal $k$ such that if we let $P=P_0$ and $P' = P_k$, there exist data sets $P_1, P_2, \cdots, P_{k-1}$ where for all $i=0,\cdots,k-1$, $P_i$ and $P_{i+1}$ are add/delete neighbors.
\end{definition}

\begin{lemma}\label{composition} \cite{DworkMNS06}
Let $X, Y$ and $Z$ be random variables. For any $\eps,{\eps}'>0$, if $X$ and $Y$ are $\eps$-indistinguishable, and $Y$ and $Z$ are ${\eps}'$-indistinguishable, then $X$ and $Z$ are $\eps+{\eps}'$-indistinguishable.
\end{lemma}

Many of our proofs use couplings so let us briefly describe on the main method we will use to construct a coupling of two random variables. Let $X$ be a random variable taking values in $\Omega_X$ and $Y$ be a random variable taking values in $\Omega_Y$. Suppose there exists a (possibly randomised) transformation $f:\Omega_X\to\Omega_Y$ such that $Y=f(X)$. That is, for all $y\in\Omega_Y$, \[\Pr(Y=y)=\sum_{x\in\Omega_X}\Pr(X=x)\Pr(f(x)=y).\] Then we can construct a coupling of $X$ and $Y$ by $\mu(x,y)=\Pr(X=x)\Pr(f(x)=y)$. A short calculation confirms that this defines a coupling. Further, notice that $\mu(x,y)\neq0$ if and only $\Pr(f(x)=y)\neq 0$. 

\begin{lemma}\label{justdistance}
Let $X$ and $Y$ be random variables taking values in $\datauni^*$ such that there exists a coupling $\mu$ such that if $\mu(x,y)\neq 0$ then the LCS distance between $x$ and $y$ is at most $A$. Then if $\mathcal{M}$ is $\eps$-deletion DP then $\mathcal{M}(X)$ and $\mathcal{M}(Y)$ are $A\eps$-indistinguishable.
\end{lemma}

\removeforcameraready{
\begin{proof} 
\begin{align*}
\Pr(\mathcal{M}(X)\in E) &= \sum_{x\in\datauni^*} \Pr(X=x)\Pr(\mathcal{M}(x)\in E)\\
&=\sum_{x\in\datauni^*} \sum_{y\in\datauni^*}\mu(x,y)\Pr(\mathcal{M}(x)\in E)\\
&\le\sum_{y\in\datauni^*} \sum_{x\in\datauni^*}\mu(x,y)e^{A\eps}\Pr(\mathcal{M}(y)\in E)\\
&=e^{A\eps} \sum_{y\in\datauni^*} \Pr(Y=y)\Pr(\mathcal{M}(y)\in E) \\
&= e^{A\eps} \Pr(\mathcal{M}(Y)\in E).
\end{align*}
\end{proof}}

\begin{lemma}[Advanced joint convexity, \cite{Balle:2018}]\label{AJC}
Let $X$ and $Y$ be random variables satisfying $X=(1-q)X_0+qX_1$ and $Y=(1-q)Y_0+qY_1$ for some $q\in[0,1]$ and random variables $X_0, X_1, Y_0$ and $Y_1$. If
\begin{itemize}
    \item $X_0$ and $Y_0$ are $\eps$-indistinguishable, 
    \item $X_1$ and $Y_1$ are $(\eps+{\eps}')$-indistinguishable and 
    \item $X_1$ and $Y_0$ are $(\eps+{\eps}')$-indistinguishable
    \item $X_0$ and $Y_1$ are $(\eps+{\eps}')$-indistinguishable
    \end{itemize}
then $X$ and $Y$ are $\left(\eps+\log(1+q(e^{{\eps}'}-1))\right)$-indistinguishable.
\end{lemma}

\removeforcameraready{\begin{proof}
\begin{align*}
    \Pr(X\in E)&=(1-q)\Pr(X_0\in E)+q\Pr(X_1\in E)\\
    &= (1-q)\Pr(X_0\in E)+qe^{-{\eps}'}\Pr(X_1\in E)\\
    &\hspace{1in}+q(1-e^{-{\eps}'})(1-q)\Pr(X_1\in E)+q(1-e^{-{\eps}'})q\Pr(X_1\in E)\\
    &\le (1-q)e^{\eps}\Pr(Y_0\in E)+qe^{\eps}\Pr(Y_1\in E)\\
    &\hspace{1in}+q(1-e^{-{\eps}'})(1-q)e^{\eps+{\eps}'}\Pr(Y_0\in E)+q(1-e^{-{\eps}'})qe^{\eps+{\eps}'}\Pr(Y_1\in E)\\
    &= (e^{\eps}+qe^{\eps}(e^{{\eps}'}-1))\Pr(Y\in E)
\end{align*}
\end{proof}}

\section{Randomized data-independent sampling} \label{sec:randomized-data-independent}

\begin{lemma}\label{changesize}
Given $m\in\mathbb{N}$, define $\mathcal{C}_m:\datauni^*\to\datauni^{m}$ be defined as follows: given a dataset $P$, form a sample $S$ by sampling $m$ data points randomly without replacement from $P$, then $\mathcal{C}_m(P)=S$. Let $P$ and $P'$ be deletion neighboring datasets and $m,m'\in\mathbb{N}$. Then if $\mathcal{M}$ is $\eps$-DP in the replacement model then $\mathcal{M}_{\mathcal{C}_{m}}(P)$ and $\mathcal{M}_{\mathcal{C}_{m'}}(P')$ are \[\left(\log\left(1+\frac{m}{|P|+1}(e^{2\eps}-1)\right)+|m-m'|\eps\right)\;\text{- indistinguishable}.\]
\end{lemma}

\begin{proof} Let $P'=P\cup\{x\}$. First, let us focus on the case where $m'=m$.
Now, 
\begin{align*}
\mathcal{M}_{\mathcal{C}_{m}}(P')&=\frac{\binom{|P|}{m}}{\binom{|P|+1}{m}}\mathcal{M}_{\mathcal{C}_{m}}(P)+\left(1-\frac{\binom{|P|}{m}}{\binom{|P|+1}{m}}\right)\mathcal{M}(\mathcal{C}_{m}(P')|_{x\in S})\\
&=\left(1-\frac{m}{|P|+1}\right)\mathcal{M}_{\mathcal{C}_{m}}(P)+\frac{m}{|P|+1}\mathcal{M}(\mathcal{C}_{m}(P')|_{x\in S}),
\end{align*}
where $\mathcal{C}_{m}(P')|_{x\in S}$ denotes the random variable $\mathcal{C}_{m}(P')$ conditioned on the event that $x\in S$.  Now, we can define a coupling of $\mathcal{C}_{m}(P)$ and $\mathcal{C}_{m}(P')|_{x\in S}$ by first sampling $S$ from $\mathcal{C}_{m}(P)$, then replacing a random element of $S$ by $x$. This coupling has LCS distance at most 2, so by Lemma~\ref{justdistance}, $\mathcal{M}_{\mathcal{C}_{m}}(P)$ and $\mathcal{M}(\mathcal{C}_{m}(P')|_{x\in S}))$ are $2\eps$-indistinguishable. Thus, by Lemma~\ref{AJC}, $\mathcal{M}_{\mathcal{C}_{m}}(P)$ and $\mathcal{M}_{\mathcal{C}_{m}}(P')$ are \[\log\left(1+\frac{m}{|P|+1}(e^{2\eps}-1)\right)\text{ -indistinguishable.}\]

Next, let us consider the case $|m-m'|=1$ and $P=P'$.
We can define a coupling of $\mathcal{C}_m(P)$ and $\mathcal{C}_{m'}(P)$ as follows: first sample $S$ from $\mathcal{C}_m(P)$, then add a random element of $P\backslash S$ to $S$. This coupling has LCS distance at most 1, so by Lemma~\ref{justdistance}, $\mathcal{M}_{\mathcal{C}_{m}}(P)$ and $\mathcal{M}_{\mathcal{C}_{m'}}(P)$ are $\eps$-indistinguishable.

Finally, we'll use Lemma~\ref{composition} to complete the proof. Note that $\mathcal{M}_{\mathcal{C}_{m}}(P)$ and $\mathcal{M}_{\mathcal{C}_{m}}(P')$ are $\log\left(1+\frac{m}{|P|+1}(e^{2\eps}-1)\right)$-indistinguishable. Then there exist $m_1, \cdots, m_{\ell-1}$ such that if we set $m_0=m$ and $m_{|m-m'|}=m'$ then for all $i$, $|m_i-m_{i-1}|\le 1$ and so $\mathcal{M}(\mathcal{C}_{m_{i-1}}(P'))$ and $\mathcal{M}(\mathcal{C}_{m_i}(P'))$ are $\eps$-indistinguishable. Therefore, by Lemma~\ref{composition}, $\mathcal{M}_{\mathcal{C}_{m}}(P)$ and $\mathcal{M}_{\mathcal{C}_{m'}}(P')$ are \[\left(\log\left(1+\frac{m}{|P|+1}(e^{2\eps}-1)\right)+|m-m'|\eps\right)\;\text{- indistinguishable}.\]
\end{proof}

\begin{definition}[log-Lipschitz functions]
A function $q: [n] \rightarrow \mathbb{R}_{\geq 0}$ is $\eps$-\emph{log-Lipschitz} if for all $m \in \{0, 1, \ldots, n-1\}$,
$
 \left \vert \log q(m+1) - \log q(m) \right \vert \leq \eps.
$
\end{definition}

\begin{lemma}
\label{lem:q-maximizer}
Let $w: [n] \rightarrow \mathbb{R}_{\geq 0}$ be nondecreasing, and let $p: [n] \rightarrow \mathbb{R}_{\geq 0}$ be any function. Then,
\[
 \max_{q:[n] \rightarrow \mathbb{R}_{\geq 0} ~\text{is}~ \eps\text{-log-Lipschitz}} \frac{\sum_{m=0}^{n} q(m) w(m) p(m)}{\sum_{m=0}^n q(m) p(m)} 
 \leq
 \frac{\sum_{m=0}^{n} e^{\eps m} w(m) p(m)}{\sum_{m=0}^n e^{\eps m} p(m)}.
\]
\end{lemma}

\begin{proof}
We will show by induction on $k = 0, 1, \ldots, n$ that we can assume w.l.o.g. that the maximizer has the form $q(0) = 1, q(1) = e^{\eps}, \ldots, q(k) = e^{\eps k}$. This holds for $k=0$ by simply normalizing. Then, assuming it holds for some $k > 0$, and given any $\eps$-log-Lipschitz $q$ such that $q(0) = 1, \ldots, q(k) = e^{\eps k}$, let us define $q'$ as follows.
\begin{align*}
    q'(m) =
    \begin{cases}
        q(m) &\text{for}~~ m = 0, 1, \ldots, k \\
        \frac{e^{\eps} q(k) q(m)}{q(k+1)} &\text{for}~~ m = k+1, \ldots, n
    \end{cases}
\end{align*}
By construction, $q'$ is $\eps$-log-Lipschitz. In particular, $q'(k+1) = e^{\eps} q(k) = e^{(k+1) \eps}$. In addition, since $q$ is $\eps$-log-Lipschitz, we have that
\begin{align*}
    \frac{e^{\eps} q(k)}{q(k+1)} \geq 1,
\end{align*}
which means that $q'(k+1) \geq q(k+1)$.

Next, we use the inequality (a slight generalization of the mediant inequality) that for $a, b, c, d > 0$ and $t \ge 1$ such that $a/b \leq c/d$,
\[
    \frac{a + c}{b + d} \leq \frac{a + tc}{b + td}
\]
Let $a = \sum_{m=0}^k q(m) w(m) p(m)$, $b = \sum_{m=0}^k q(m) p(m)$, $c = \sum_{k+1}^n q(m) w(m) p(m)$, and $d = \sum_{k+1}^n q(m) p(m)$, and $t = e^{\eps} q(k) / q(k+1)$. By the non-decreasing property of $w$, we have that
\begin{align*}
    a/b \leq w(k) \leq w(k+1) \leq c/d.
\end{align*}
Therefore, the inequality above, and by definition of $q'$, we have that
\begin{align*}
    \frac{\sum_{m=0}^n q(m) w(m) p(m)}{\sum_{m=0}^n q(m) p(m)}
    &\leq 
    \frac{\sum_{m=0}^k q(m) w(m) p(m) + e^{\eps} (q(k) / q(k+1)) \cdot \sum_{m=k+1}^n q(m) w(m) p(m)}{\sum_{m=0}^k q(m) p(m) + e^{\eps} (q(k) / q(k+1)) \cdot \sum_{m=k+1}^n q(m) p(m) } \\
    &= \frac{\sum_{m=0}^n q'(m) w(m) p(m)}{\sum_{m=0}^n q'(m) p(m)}.
\end{align*}
So, by induction, we can assume that the maximizer has the form $q(0) = 1, q(1) = e^{\eps}, \ldots, q(n) = e^{\eps n}$,
which completes the proof.
\end{proof}

\begin{proof}[Proof of Theorem~\ref{thm:random-data-independent-sampling}] Let $\mathcal{C}_m:\datauni^*\to\datauni^m$ be the sampling scheme that given a dataset $P$, returns $S$ where $S$ is a uniformly random subset of $P$ of size $m$ (drawn without replacement).
Let $y \in \mathcal{O}$ be any outcome, and let $P \sim P'$ be neighboring datasets. Then, we have that
\begin{align*}
    \Pr[\mathcal{M}_{\mathcal{C}}(P) = y]
    &= \sum_{m=0}^n \Pr [\mathcal{M}_{\mathcal{C}_m}(P) = y] \cdot \Pr[|\mathcal{C}(P)| = m] \\
    &\leq \sum_{m=0}^n \left(1 + \frac{m}{n} (e^{\eps} - 1) \right) \cdot Pr [\mathcal{M}_{\mathcal{C}_m}(P') = y ] \cdot \numsamples(m) \\
    &\leq \frac{\sum_{m=0}^n \left(1 + \frac{m}{n}(e^\eps - 1) \right) \cdot e^{\eps m} \cdot \numsamples(m)}{\sum_{m=0}^n e^{\eps m} \numsamples(m)}\cdot \sum_{m=0}^n \Pr[\mathcal{M}_{\mathcal{C}_m}(P') = y  ] \cdot \numsamples(m) \\
    &= \left( 1 + \frac{\E_{m \sim \tilde{\numsamples}}[m]}{n} (e^{\eps} - 1) \right) \cdot \Pr[\mathcal{M}_{\mathcal{C}}(P') = y]
\end{align*}
where the first inequality follows from Lemma~\ref{changesize}. Then, note that $(1 + (m/n) (e^{\eps} - 1))$ is non-decreasing, and that $\Pr [\mathcal{M}_{\mathcal{C}_m}(P')) = y ]$ is $\eps$-log-Lipschitz by definition, so the second inequality follows by Lemma~\ref{lem:q-maximizer}. After rearranging and simplifying, we obtain the desired result.

Finally, for the lower bound, suppose the data universe $\mathcal{U}=[0,1]$.
Let $P = \{1,\cdots,1\}$ consist of $n$ 1s and $P'$ be the neighboring dataset $P'=P\backslash\{1\}\cup\{0\}$. Let $\mathcal{M}:\mathcal{U}^*\to\mathbb{R}$ be defined by $\mathcal{M}(S) = \sum_{x\in S} \indicator{x=1}+\Lap\left(1/\eps\right)$ so $\mathcal{M}$ is $\eps$-deletion DP. Then 
\begin{align*}
\frac{\Pr(\mathcal{M}_{\mathcal{C}}(P')=n)}{\Pr(\mathcal{M}_{\mathcal{C}}(P)=n)} &= \frac{\sum_{m=0}^n \Pr(\numsamples = m)\left(\frac{m}{n}e^{-(n-m+1)\eps}+\left(1-\frac{m}{n}\right)e^{-(n-m)\eps}\right)}{\sum_{m=0}^n \Pr(\numsamples = m)e^{-(n-m)\eps}}\\
\removeforcameraready{
&= \frac{\sum_{m=0}^n \Pr(\numsamples = m)\left(\frac{m}{n}e^{(m-1)\eps}+\left(1-\frac{m}{n}\right)e^{m\eps}\right)}{\sum_{m=0}^n \Pr(\numsamples = m)e^{m\eps}}\\
&= \frac{\sum_{m=0}^n \Pr(\numsamples = m)e^{m\eps}\left(1-\frac{m}{n}(1-e^{-\eps})\right)}{\sum_{m=0}^n \Pr(\numsamples = m)e^{m\eps}}\\}
&= 1-\frac{1}{n}(1-e^{-\eps})\frac{\sum_{m=0}^n \Pr(\numsamples = m)e^{m\eps} m}{\sum_{m=0}^n \Pr(\numsamples = m)e^{m\eps}}.
\end{align*}
Thus, taking the reciprocal,
\[\log\frac{\Pr(\mathcal{M}_{\mathcal{C}}(P)=n)}{\Pr(\mathcal{M}_{\mathcal{C}}(P')=n)} = -\log\left(1-\frac{1}{n}(1-e^{-\eps})\frac{\sum_{m=0}^n \Pr(\numsamples = m)e^{m\eps} m}{\sum_{m=0}^n \Pr(\numsamples = m)e^{m\eps}}\right).\]
\end{proof}

\section{Data-dependent sampling} \label{sec:data-dependent}

\begin{proof}[Proof of Proposition~\ref{advantage}: hypothesis testing perspective]
Let $H:\mathbb{N}\to\{0,1\}$ be the hypothesis test such that for all $x\in\mathbb{N}$, and $b\in\{0,1\}$, $e^{-\eps}\Pr(H(x)=b)\le \Pr(H(x+1)=b)\le e^{\eps}\Pr(H(x)=b)$. Then $H':\datauni^*\to\{0,1\}$ defined by $H'(S)=H(|S|)$ is $\eps$-deletion DP. By assumption, $H'_{\mathcal{C}_{\tilde{f}}}$ is $\eps'$-DP. This implies that $H(\tilde{f}(P))$ and $H(\tilde{f}(P'))$ are $\eps'$-indistinguishable. Therefore, \[\adv(H) = \Pr[H(\tilde{f}(P)) = 0] - \Pr[H(\tilde{f}(P')) = 0] \leq \Pr[H(\tilde{f}(P')) = 0](e^{\eps'} - 1) \leq e^{\eps'} - 1.\] The result follows from taking the supremum over all $\eps$-DP $H$.
\end{proof}

\begin{proof}[Proof of Theorem~\ref{onestratumrr}: proportional allocation with randomized rounding]
Let $P$ be a dataset, $x$ be a data point and $P'=P\cup\{x\}$. Let $m=r|P|$, $m'=r|P'|$, $m^L=\lfloor m\rfloor$, ${m'}^L=\lfloor m'\rfloor$, $p=m-m^L$ and $p'=m'-{m'}^L$. 
Now, $m'-m=r<1$ so we have two cases, $m^L={m'}^L$ or $m^L={m'}^L-1$.

As in Lemma~\ref{changesize}, let $\mathcal{C}_m:\datauni^*\to\datauni^m$ be the sampling scheme that given a dataset $P$, returns $S$ where $S$ is a uniformly random subset of $P$ of size $m$ (drawn without replacement).
Note that by Theorem~\ref{priorwork}, for $m,m'\in\mathbb{N}$, $\mathcal{M}_{m}(P)$ and $\mathcal{M}_{m'}(P)$ are $(|m-m'|\eps)$-indistinguishable, and $\mathcal{M}_{\mathcal{C}_m}(P)$ and $\mathcal{M}_{\mathcal{C}_{m'}}(P')$ are $\left(\log\left(1+\frac{m}{|P|+1}(e^{2\eps}-1)\right)+|m-m'|\eps\right)$-indistinguishable. 

Firstly, suppose $m^L={m'}^L$. Let
\begin{align*}
    \mu_0 &=\frac{1}{1-r}((1-p-r)\mathcal{M}_{\mathcal{C}_{m^L}}(P)+p\mathcal{M}_{\mathcal{C}_{m^L+1}}(P)), \\ 
    \mu_0' &= \frac{1}{1-r}((1-p-r)\mathcal{M}_{\mathcal{C}_{m^L}}(P')+p\mathcal{M}_{\mathcal{C}_{m^L+1}}(P')), \\ \mu_1 &=\mathcal{M}_{\mathcal{C}_{m^L}}(P), \\
    \mu_1' &=\mathcal{M}_{\mathcal{C}_{m^L+1}}(P').
\end{align*}
Notice that $\mathcal{M}_{\mathcal{C}_{r}}(P)=(1-r)\mu_0+r\mu_1$ and $\mathcal{M}_{\mathcal{C}_{r}}(P')=(1-r)\mu_0'+r\mu_1'$. Now, by Lemma~\ref{AJC} and Lemma~\ref{justdistance}, $\mu_0$ and $\mu_0'$ are $\log(1+\frac{m^L+1}{|P|+1}(e^{2\eps}-1))$-indistinguishable. Further, all the pairs $(\mu_0',\mu_1)$, $(\mu_1,\mu_1')$ and $(\mu_0,\mu_1')$ are $\left(\log(1+\frac{m^L+1}{|P|+1}(e^{2\eps}-1))+\eps\right)$-indistinguishable. Therefore, by Lemma~\ref{AJC}, $\mathcal{M}_{\mathcal{C}_{r}}(P)$ and $\mathcal{M}_{\mathcal{C}_{r}}(P')$ are ${\eps}'$-indistinguishable where 
\begin{align*}
    {\eps}' &\le \log\left(1+\frac{m^L+1}{|P|+1}(e^{2\eps}-1)\right)+\log(1+r(e^{\eps}-1)) \\
    &\le \log\left(1+\left(r+\frac{1}{|P|+1}\right)(e^{2\eps}-1)\right)+\log(1+r(e^{\eps}-1)).
\end{align*}
Next, suppose ${m'}^L=m^L+1$. Let $1-q=\min\{p,1-p'\}$ and 
\begin{align*}
    \mu_0 &=\mathcal{M}_{\mathcal{C}_{m^L+1}}(P), \\
    \mu_0' &=\mathcal{M}_{\mathcal{C}_{m^L+1}}(P'), \\
    \mu_1 &=\frac{1}{q}((p-1+q)\mathcal{M}_{\mathcal{C}_{m^L+1}}(P)+(1-p)\mathcal{M}_{\mathcal{C}_{m^L}}(P)), \\
    \mu_1' &=\frac{1}{q}((1-p'-1+q)\mathcal{M}_{\mathcal{C}_{m^L+1}}(P')+p'\mathcal{M}_{\mathcal{C}_{m^L+2}}(P')).
\end{align*}
Notice that 
\begin{align*}
\mathcal{M}_{\mathcal{C}_{r}}(P)=(1-q)\mu_0+q\mu_1 ~~~~~~\text{and}~~~~~~
\mathcal{M}_{\mathcal{C}_{r}}(P')=(1-q)\mu_0'+q\mu_1'.
\end{align*}
Now, by Lemma~\ref{priorwork}, $\mu_0$ and $\mu_0'$ are $\log\left(1+\frac{m^L+1}{|P|+1}\right)$-indistinguishable. Further, all the pairs $(\mu_0',\mu_1)$, $(\mu_1,\mu_1')$ and $(\mu_0,\mu_1')$ are $\left(\log(1+\frac{m^L+1}{|P|+1}(e^{2\eps}-1))+2\eps\right)$-indistinguishable. Also, note that $q\le r$. Then by Lemma~\ref{AJC}, $\mathcal{M}_{\mathcal{C}_{r}}(P)$ and $\mathcal{M}_{\mathcal{C}_{r}}(P')$ are ${\eps}'$-indistinguishable where 
\begin{align*}
    {\eps}' &\le \log\left(1+\frac{m^L+1}{|P|+1}(e^{2\eps}-1)\right)+\log(1+p(e^{2\eps}-1)) \\
    &\le \log\left(1+\left(r+\frac{1}{|P|+1}\right)(e^{2\eps}-1)\right)+\log(1+r(e^{2\eps}-1)).
\end{align*}
\end{proof}

\section{Cluster sampling} \label{app:cluster}
\begin{proof}[Proof of Theorem~\ref{clusterUB}] 
Without loss of generality, let $i=1$. Notice that conditioned on cluster $1\notin I$, the distribution of outputs of $\mathcal{M}_{\mathcal{C}}(P)$ and $\mathcal{M}_{\mathcal{C}}(P')$ are identical. Let $E$ be a set of outcomes. Then
\begin{align*}
\Pr(\mathcal{M}_{\mathcal{C}}(P)\in E) &= \frac{\clusterrate}{k}\Pr(\mathcal{M}_{\mathcal{C}}(P)\in E\;|\; 1\in I)+\left(1-\frac{\clusterrate}{k}\right)\Pr(\mathcal{M}_{\mathcal{C}}(P)\in E\;|\; 1\notin I)\\
&= \frac{\clusterrate}{k}\Pr(\mathcal{M}_{\mathcal{C}}(P)\in E\;|\; 1\in I)+\left(1-\frac{\clusterrate}{k}\right)\Pr(\mathcal{M}_{\mathcal{C}}(P')\in E\;|\; 1\notin I).
\end{align*}
Now, we have that
\begin{align*}
    \frac{\clusterrate}{k}\Pr(\mathcal{M}_{\mathcal{C}}(P)\in E\;|\; 1\in I) 
    &= \frac{\clusterrate}{k} \sum_{|I|=\clusterrate, 1\in I} \frac{1}{\binom{k}{\clusterrate}} \Pr(\mathcal{M}(P_I)\in E)\\
    &\le \frac{\clusterrate}{k} \sum_{|I|=\clusterrate, 1\in I} \frac{1}{\binom{k}{\clusterrate}} e^{\eps}\Pr(\mathcal{M}(P'_I)\in E)\\
    &= \frac{\clusterrate}{k}e^{\eps}\Pr(\mathcal{M}_{\mathcal{C}}(P')\in E\;|\; 1\in I),
\end{align*}
where the inequality follows from the fact that the LCS distance between $P_I$ and $P'_I$ is 1.
Thus,
\begin{align*}
    \Pr(\mathcal{M}_{\mathcal{C}}(P)\in E)&\le \frac{\clusterrate}{k}e^{\eps}\Pr(\mathcal{M}_{\mathcal{C}}(P')\in E\;|\; 1\in I)+\left(1-\frac{\clusterrate}{k}\right)\Pr(\mathcal{M}_{\mathcal{C}}(P')\in E\;|\; 1\notin I)\\
    &= \Pr(\mathcal{M}_{\mathcal{C}}(P')\in E)+\frac{\clusterrate}{k}(e^{\eps}-1)\Pr(\mathcal{M}_{\mathcal{C}}(P')\in E\;|\; 1\in I).
\end{align*}
Now, we need to relate $\Pr(\mathcal{M}_{\mathcal{C}}(P')\in E\;|\; 1\in I)$ to $\Pr(\mathcal{M}_{\mathcal{C}}(P)\in E)$.
For a set $I$ such that $1\notin I$ and index $i\in I$, let $I\cup\{1\}\backslash\{i\}$ be the set where index $i$ has been replaced with 1. 
Then, 
\begin{align*}
    \left(1-\frac{\clusterrate}{k}\right)\Pr(\mathcal{M}_{\mathcal{C}}(P')\in E\;|\; 1\notin I) &= \sum_{|I|=\clusterrate, 1\notin I} \frac{1}{\binom{k}{\clusterrate}} \Pr(\mathcal{M}(P_I)\in E)\\
    &= \sum_{|I|=\clusterrate, 1\notin I} \sum_{i\in I} \frac{1}{\clusterrate} \frac{1}{\binom{k}{\clusterrate}} \Pr(\mathcal{M}(P_I)\in E)\\
    &\ge \sum_{|I|=\clusterrate, 1\notin I} \sum_{i\in I} \frac{1}{\clusterrate} \frac{1}{\binom{k}{\clusterrate}} e^{-(n_1+n_i)\eps}\Pr(\mathcal{M}(P_{I\cup\{1\}\backslash\{i\}})\in E)\\
    &\ge e^{-(n_1+n_{\max})\eps} \frac{1}{\clusterrate}\sum_{|I|=\clusterrate, 1\notin I} \sum_{i\in I} \frac{1}{\binom{k}{\clusterrate}} \Pr(\mathcal{M}(P_{I\cup\{1\}\backslash\{i\}})\in E),
\end{align*}
where the first inequality follows from the fact that the LCS distance between $P_I$ and $P_{I\cup\{1\}\backslash\{i\}}$ is at most $n_1+n_i$. Now, notice that the sets $I\cup\{1\}\backslash\{i\}$ in the above sum all contain 1, and each index $I'$ such that $|I'|=\clusterrate$ and $1\in I'$ appears in the sum $k-\clusterrate$ times (corresponding to the $k-\clusterrate$ possible choices for the swapped index $i$). Therefore, we can rewrite the sum as
\begin{align*}
    \left(1-\frac{\clusterrate}{k}\right)\Pr(\mathcal{M}_{\mathcal{C}}(P')\in E\;|\; 1\notin I)
    &\ge e^{-(n_1+n_{\max})\eps}\frac{k-\clusterrate}{\clusterrate} \sum_{|I|=\clusterrate, 1\in I} \frac{1}{\binom{k}{\clusterrate}} \Pr(\mathcal{M}(P_I)\in E)\\
    \removeforcameraready{
    &= e^{-(n_1+n_{\max})\eps}\frac{k-\clusterrate}{\clusterrate}\frac{\clusterrate}{k}\Pr(\mathcal{M}_{\mathcal{C}}(P')\in E\;|\; 1\in I)\\}
    & = e^{-(n_1+n_{\max})\eps}\left(1-\frac{\clusterrate}{k}\right)\Pr(\mathcal{M}_{\mathcal{C}}(P')\in E\;|\; 1\in I).
\end{align*}
Thus, 
\begin{align*}
    \Pr(\mathcal{M}_{\mathcal{C}}(P')\in E)&= \frac{\clusterrate}{k}\Pr(\mathcal{M}_{\mathcal{C}}(P')\in E\;|\; 1\in I)+\left(1-\frac{\clusterrate}{k}\right)\Pr(\mathcal{M}_{\mathcal{C}}(P')\in E\;|\; 1\notin I)\\
    &\ge \frac{\clusterrate}{k}\Pr(\mathcal{M}_{\mathcal{C}}(P')\in E\;|\; 1\in I)+\left(1-\frac{\clusterrate}{k}\right)e^{-(n_1+n_{\max})\eps}\Pr(\mathcal{M}_{\mathcal{C}}(P')\in E\;|\; 1\in I)\\
    &= \left(\frac{\clusterrate}{k}+\left(1-\frac{\clusterrate}{k}\right)e^{-(n_1+n_{\max})\eps}\right)\Pr(\mathcal{M}_{\mathcal{C}}(P')\in E\;|\; 1\in I).
\end{align*} Finally, 
\begin{align*}
\Pr(\mathcal{M}_{\mathcal{C}}(P)\in E) 
&\le \Pr(\mathcal{M}_{\mathcal{C}}(P')\in E)+\frac{\clusterrate}{k}(e^{\eps}-1)\Pr(\mathcal{M}_{\mathcal{C}}(P')\in E\;|\; 1\in I)\\
&\le \Pr(\mathcal{M}_{\mathcal{C}}(P')\in E)+\frac{\clusterrate}{k}(e^{\eps}-1)\frac{1}{\left(\frac{\clusterrate}{k}+\left(1-\frac{\clusterrate}{k}\right)e^{-(n_1+n_{\max})\eps}\right)}\Pr(\mathcal{M}_{\mathcal{C}}(P')\in E)\\
&\le \left(1+\frac{\clusterrate}{k}(e^{\eps}-1)\frac{1}{\left(\frac{\clusterrate}{k}+\left(1-\frac{\clusterrate}{k}\right)e^{-(n_1+n_{\max})\eps}\right)}\right)\Pr(\mathcal{M}_{\mathcal{C}}(P')\in E)
\end{align*}
\end{proof}
\removeforcameraready{Now we turn our attention to the lower bound.}
\begin{proof}[Proof of Theorem~\ref{clusterLB}]
Let $C_1=\{1,\cdots,1\}$ and $C_j=\{-1,\cdots,-1\}$ for all $j\in\{2,\cdots,k\}$. Let $C_1'=C_1\backslash\{1\}\cup\{-1\}$ be the same as $C_1$ except with one 1 switched to a $-1$. Let $\mathcal{M}(S)=\sum_{x\in S}~x~+~\Lap(1/\eps),$ so $\mathcal{M}$ is $\eps$-deletion DP. Notice that $\mathcal{M}$ has the property that if $\sum_{x\in S'} x=\sum_{x\in S} x+a$, for some $a\in\mathbb{R}$ then $\Pr(\mathcal{M}(S)=\sum_{x\in S} x)=e^{|a|\eps} \Pr(\mathcal{M}(S')=\sum_{x\in S} x)$. This equality allows us to tighten many of the inequalities that appeared in the proof of Theorem~\ref{clusterUB}, and give a lower bound.
\begin{align*}
    \Pr\left(\mathcal{M}_{\mathcal{C}}(P)=n_1+1\right)&=\frac{\clusterrate}{k}\Pr\left(\mathcal{M}_{\mathcal{C}}(P)=n_1+1\;|\; 1\in I\right)+\left(1-\frac{\clusterrate}{k}\right)\Pr\left(\mathcal{M}_{\mathcal{C}}(P)=n_1+1\;|\; 1\notin I\right)\\
    &=\frac{\clusterrate}{k}e^{\eps}\Pr\left(\mathcal{M}_{\mathcal{C}}(P')=n_1+1\;|\; 1\in I\right)+\left(1-\frac{\clusterrate}{k}\right)\Pr\left(\mathcal{M}_{\mathcal{C}}(P')=n_1+1\;|\; 1\notin I\right)\\
    &= \Pr\left(\mathcal{M}_{\mathcal{C}}(P')=n_1+1\right)+\frac{\clusterrate}{k}(e^{\eps}-1)\Pr\left(\mathcal{M}_{\mathcal{C}}(P')=n_1+1\;|\; 1\in I\right).
\end{align*}
Now, 
\begin{align*}
    &\left(1-\frac{\clusterrate}{k}\right)\Pr(\mathcal{M}_{\mathcal{C}}(P')\in E\;|\; 1\notin I) = \sum_{|I|=\clusterrate, 1\notin I} \frac{1}{\binom{k}{\clusterrate}} \Pr(\mathcal{M}(P_I)\in E)\\
    &= \sum_{|I|=\clusterrate, 1\notin I} \sum_{i\in I} \frac{1}{\clusterrate} \frac{1}{\binom{k}{\clusterrate}} \Pr(\mathcal{M}(P_I)\in E)
    = \sum_{|I|=\clusterrate, 1\notin I} \sum_{i\in I} \frac{1}{\clusterrate} \frac{1}{\binom{k}{\clusterrate}} e^{-(n_1+n_i)\eps}\Pr(\mathcal{M}(P_{I\cup\{1\}\backslash\{i\}})\in E)\\
    &\le e^{-(n_1+n_{\min})\eps} \frac{1}{\clusterrate}\sum_{|I|=\clusterrate, 1\notin I} \sum_{i\in I} \frac{1}{\binom{k}{\clusterrate}} \Pr(\mathcal{M}(P_{I\cup\{1\}\backslash\{i\}})\in E)\\
    &= e^{-(n_1+n_{\min})\eps}\frac{k-\clusterrate}{\clusterrate} \sum_{|I|=\clusterrate, 1\in I} \frac{1}{\binom{k}{\clusterrate}} \Pr(\mathcal{M}(P_I)\in E) \removeforcameraready{\\
    &} = e^{-(n_1+n_{\min})\eps}\left(1-\frac{\clusterrate}{k}\right)\Pr(\mathcal{M}_{\mathcal{C}}(P')\in E\;|\; 1\in I).
\end{align*}
Thus, 
\begin{align*}
    \Pr(\mathcal{M}_{\mathcal{C}}(P')\in E)&= \frac{\clusterrate}{k}\Pr(\mathcal{M}_{\mathcal{C}}(P')\in E\;|\; 1\in I)+\left(1-\frac{\clusterrate}{k}\right)\Pr(\mathcal{M}_{\mathcal{C}}(P')\in E\;|\; 1\notin I)\\
    &\le \frac{\clusterrate}{k}\Pr(\mathcal{M}_{\mathcal{C}}(P')\in E\;|\; 1\in I)+\left(1-\frac{\clusterrate}{k}\right)e^{-(n_1+n_{\min})\eps}\Pr(\mathcal{M}_{\mathcal{C}}(P')\in E\;|\; 1\in I)\\
    &= \left(\frac{\clusterrate}{k}+\left(1-\frac{\clusterrate}{k}\right)e^{-(n_1+n_{\min})\eps}\right)\Pr(\mathcal{M}_{\mathcal{C}}(P')\in E\;|\; 1\in I).
\end{align*} Finally, 
\begin{align*}
\Pr(\mathcal{M}_{\mathcal{C}}(P)\in E) 
&= \Pr(\mathcal{M}_{\mathcal{C}}(P')\in E)+\frac{\clusterrate}{k}(e^{\eps}-1)\Pr(\mathcal{M}_{\mathcal{C}}(P')\in E\;|\; 1\in I)\\
&\ge \Pr(\mathcal{M}_{\mathcal{C}}(P')\in E)+\frac{\clusterrate}{k}(e^{\eps}-1)\frac{1}{\left(\frac{\clusterrate}{k}+\left(1-\frac{\clusterrate}{k}\right)e^{-(n_1+n_{\min})\eps}\right)}\Pr(\mathcal{M}_{\mathcal{C}}(P')\in E)\\
&= \left(1+\frac{\clusterrate}{k}(e^{\eps}-1)\frac{1}{\left(\frac{\clusterrate}{k}+\left(1-\frac{\clusterrate}{k}\right)e^{-(n_1+n_{\min})\eps}\right)}\right)\Pr(\mathcal{M}_{\mathcal{C}}(P')\in E).
\end{align*}
\end{proof}

\begin{proof}[Proof of Theorem~\ref{clusterHT}]
Let $D = \mathcal{C}_{\clusterrate}(P)$ and $D' = \mathcal{C}_{\clusterrate}(P')$. For an event $E \in \outputspace$, define the probabilities $p, q, p'$ and $q'$ as follows.
\begin{align*}
    p = \Pr( \mathcal{M}(D) \in E \vert C_1 \in D) ~~~~&~~~ 
    q = \Pr(\mathcal{M}(D) \in E | C_1 \notin D) \\
    p' = \Pr(\mathcal{M}(D') \in E | C_1 \in D') ~~~~&~~~
    q' = \Pr(\mathcal{M}(D') \in E | C_1 \notin D')
\end{align*}
By the existence of $\mathcal{H}$ described in the lemma statement, there must exist an event $E$ such that $q \le e^{-\eps'}p$. 
Since $P$ and $P'$ only differ on $C_1$, the distributions of $\mathcal{M}(D)|_{C_1 \notin D}$ and $\mathcal{M}(D')|_{C_1 \notin D'}$ are identical, which means that $q = q'$. 
Then, we can compute a lower bound on the indistinguishability of $\mathcal{M}(D)$ and $\mathcal{M}(D')$ as follows. Without loss of generality, assume $p'>p$, and proceed as follows.
\begin{align*}
    \frac{\Pr( \mathcal{M}(D') \in E) }{\Pr (\mathcal{M}(D) \in E )} 
    &= \frac{p' \cdot Pr(C_1 \in D') + q \cdot Pr(C_1 \notin D')}{p \cdot Pr(C_1 \in D) + q \cdot Pr(C_1 \notin D)} \\
    &= \frac{p' \cdot \frac{\clusterrate}{k} + q \cdot (1- \frac{\clusterrate}{k}) }{p \cdot \frac{\clusterrate}{k} + q \cdot (1- \frac{\clusterrate}{k})} \\
   \removeforcameraready{ &= \frac{p \cdot \frac{\clusterrate}{k} + q \cdot (1-\frac{\clusterrate}{k}) + (p'-p) \cdot \frac{\clusterrate}{k}}{p \cdot \frac{\clusterrate}{k} + q \cdot (1- \frac{\clusterrate}{k})} \\
    &= 1 + \frac{ (p'-p) \frac{\clusterrate}{k}}{p \cdot \frac{\clusterrate}{k} + q \cdot (1- \frac{\clusterrate}{k})} \\}
    &\geq 1 + \frac{ (p'-p) \frac{\clusterrate}{k}}{p \cdot (\frac{\clusterrate}{k} + e^{-\eps'} (1- \frac{\clusterrate}{k}))} \\ 
    &= 1 + \left( \frac{p'}{p}-1 \right) \frac{\frac{\clusterrate}{k}}{\frac{\clusterrate}{k} + e^{-\eps'} (1-\frac{\clusterrate}{k})}
\end{align*}
where the final inequality follows from the fact that $\mathcal{M}$ is $\epsilon$-DP, so $p'/p\ge e^{\epsilon''}$ by definition.
\end{proof}

\section{Stratified sampling}

\begin{proof}[Proof of Theorem~\ref{sswor}: proportional allocation for stratified sampling] Given $\mathcal{M}:([k]\times\datauni)^*\to\mathcal{Y}$,
for all datasets $T_2, \cdots, T_k\in\datauni^*$, define $\mathcal{M}^{T_2,\cdots,T_k}:\datauni^*\to\mathcal{Y}$ by $\mathcal{M}^{T_2,\cdots,T_k}(S)=\mathcal{M}(S\sqcup T_2\sqcup\cdots\sqcup T_k)$. Then since $\mathcal{M}$ was $(\eps,\cdots,\eps)$-stratified deletion DP, ${\mathcal{M}^{T_2,\cdots,T_k}}$ is $\eps$-deletion DP. Let $\mathcal{C}_r$ be as in Lemma~\ref{onestratumrr} so for all $S,S'$ add/delete neighbours such that $r|S|\ge 1$ and $r|S'|\ge 1$, $\mathcal{M}^{T_2,\cdots,T_k}_{\mathcal{C}_r}(S)$ and $\mathcal{M}^{T_2,\cdots,T_k}_{\mathcal{C}_r}(S')$ are $\eps'$-indistinguishable where \[\eps'\le\log\left(1+2r(e^{2\eps}-1)\right)+\log(1+r(e^{2\eps}-1)).\]
Now, let $P=S_1\sqcup S_2\sqcup\cdots\sqcup S_k$ and $P=S_1'\sqcup S_2\sqcup\cdots\sqcup S_k$ be deletion stratified neighboring datasets that differ in the first stratum. Since $S_2\sqcup\cdots\sqcup S_k$ are shared between $P$ and $P'$, and the datasets $T_i$ only dependent on strata $S_i$, the distribution of $T_2, \cdots, T_k$ are identical given inputs $P$ and $P'$. Let $q$ be the distribution of $T_2,\cdots,T_k$ so $q(T_2,\cdots,T_k) = \Pr(\mathcal{C}_r(S_2)=T_2, \cdots, \mathcal{C}_r(S_k)=T_k)$. Then given an event $E$,
\begin{align*}
\Pr(\mathcal{M}_{\mathcal{C}_{\bm{f}_{{\rm prop}, r}}}(P)\in E) &= \int_{T_2, \cdots, T_k} q(T_2, \cdots, T_k)\Pr(\mathcal{M}^{T_2,\cdots, T_k}_{\mathcal{C}_r}(S_1)\in E)\\
&\le \int_{T_2, \cdots, T_k} q(T_2, \cdots, T_k)e^{\eps'}\Pr(\mathcal{M}^{T_2,\cdots, T_k}_{\mathcal{C}_r}(S_1')\in E)\\
&= e^{\eps'}\Pr(\mathcal{M}_{\mathcal{C}_{\bm{f}_{{\rm prop}, r}}}(P')\in E).
\end{align*}
\end{proof}

\subsection{Neyman Allocation on Simulated Data}
\label{app:neyman-CBP}

\textbf{Simulating Population Data.} 
We start with County Business Patterns (CBP) data, which is published by the U.S. Census Bureau~\cite{eckert2021}. The released data is a tabulated version of
the underlying microdata from the Business Register (BR), a database of all known single and multi-establishment employer companies. 

The CBP data contains the total number of establishments and the total number of employees by industry, down to the county level. For each industry $\times$ county, the data also contains information on how many establishments fall into different Employee Size Classes (1-4 employees, 5-9 employees, etc.).\footnote{The codebook for this dataset can be found here: \url{https://www2.census.gov/programs-surveys/cbp/technical-documentation/records-layouts/2015\_record_layouts/county\_layout\_2015.txt}.} 
In order to bound the sensitivity of the allocation function we will apply to this data,
we top-code the Employee Size at 10,000 employees. 

Using the CBP data, we generate a simulated data set that is consistent with the tabulated release to serve as our population data. Each row of the simulated data set corresponds to an establishment with attributes of industry, county, and employee size. Through linear programming, we generate these simulated establishments such that the counts within the Employee Size Classes, total number of establishments, and total number of employees by county and industry remain consistent with the CBP tabulations. 
The establishments are stratified by Employee Size into $k=12$ strata.\footnote{The strata are as follows: $[1,4], [5,9], [10,19], [20,49], [50,99], [100,249], [250,499], [500, 999], [1000, 1499], [1500, 2499],$ \\
$[2500, 4999], [5000, 10000]$}.

\noindent \textbf{Neyman allocation.} We apply Neyman allocation with a target final sample size of $m=10,000$. 
We can find two neighboring populations (the first population was obtained from the CBP data, and the neighbouring population was obtained by taking the company with the fewest employees and increasing their employee count to the maximum of 10,000) such that the Neyman allocation for one population is $[1261, 621, 517, 1969, 833, 1947, 1058, 762, 257, 248, 306, 225]$, and the Neyman allocation for the second population is $[1259, 620, 516, 1965, 831, 1943, 1056, 761, 257, 247, 306, 244]$. While these allocations are fairly similar, they do differ by 19 samples in the last stratum that corresponds to establishments with Employee Sizes of 10,000 or above. The impact of this difference in the last stratum depends on the target statistic: it might not have a large impact on the weighted mean of Employee Size, but could lead to more substantial changes for other statistics. As an illustrative example, we can consider the target statistic of simply estimating the stratum sizes of the sample (ie. the allocation itself) in a privacy-preserving manner. For such a goal, these differences in the last stratum would lead to significant privacy degradation. Note that such a change is a large enough to be detectable by an $\eps$-DP mechanism for moderate values of $\eps$ ($\eps>0.1$ is sufficient).

\section{Instantiating the estimation-based lower bound} \label{app:estimation-lb}

For any function $f$, the accuracy to which one can privately estimate $f$ depends on the \emph{sensitivity} of $f$, i.e., how much changing the input dataset can change the output.

\begin{proof}[Proof of Proposition~\ref{var-LB}]
Define $\mathcal{M}_{SS}:\datauni^*\to\mathbb{N}$ as follows. For all $P\in\datauni^*$, $\mathcal{M}(P)=|P|+\Lap(1/\eps).$ Then $\mathcal{M}$ is $\eps$-deletion DP.
Suppose that $\tilde{f}:\datauni^*\to\mathbb{N}$ is such that for all $\eps$-deletion DP mechanisms $\mathcal{A}$, $\mathcal{A}_{\mathcal{C}_{\tilde{f}}}$ is $\eps'$-replacement DP. This implies that $\mathcal{M}_{\mathcal{C}_{\tilde{f}}}(P))=\tilde{f}(P)+\Lap(1/\eps)$
is $\eps'$-replacement DP. Therefore, by the definition of $\alpha$, there exists a population $P$ such that
$\sup_{P\in\datauni^n} \mathbb{E}[|\mathcal{M}_{\mathcal{C}_{\tilde{f}}}(P))-f(P)|^2] \ge \alpha.$
Also 
\begin{align*}
\alpha\le \mathbb{E}[|\mathcal{M}_{\mathcal{C}_{\tilde{f}}}(P))-f(P)|^2]&=\mathbb{E}[|\tilde{f}(P)+\Lap(1/{\eps})-f(P)|^2]\removeforcameraready{\\
&= \mathbb{E} [(\tilde{f}(P)-f(P))^2+\Lap(1/{ \eps})(\tilde{f}(P)-f(P))+(\Lap(1/{\eps}))^2]\\
&}= \mathbb{E}[]|\tilde{f}(P)-f(P)|^2]+(1/{\eps})^2.
\end{align*}
After a small amount of rearranging we arrive at the result. 
\end{proof}

\end{document}